\documentclass{article}

\usepackage{arxiv}
\usepackage{wrapfig}
\usepackage{amsmath}

\usepackage{color}
\usepackage{graphicx,subfigure}
\usepackage{xspace}
\usepackage{eso-pic}

\usepackage[utf8]{inputenc} 
\usepackage[T1]{fontenc}    
\usepackage{hyperref}       
\usepackage{url}            
\usepackage{booktabs}       
\usepackage{amsfonts}       
\usepackage{nicefrac}       
\usepackage{microtype}      
\usepackage{lipsum}
\title{How Can CNNs Use Image Position for Segmentation?}

\author{
 Rito Murase \\
  Tohoku University\\
  \texttt{rmurase@vision.is.tohoku.ac.jp} \\
   \And
 Masanori Suganuma \\
  Tohoku University \\
  RIKEN AIP \\
  \texttt{suganuma@vision.is.tohoku.ac.jp} \\
  \And
 Takayuki Okatani \\
  Tohoku University \\
  RIKEN AIP \\ 
  \texttt{okatani@vision.is.tohoku.ac.jp} \\
}

\def\bmvaHangBox#1{\begin{minipage}[t]{\textwidth}\begin{tabbing}~\\[-\baselineskip]#1\end{tabbing}\end{minipage}}

\begin{document}

\maketitle

\begin{abstract}
Convolution is an equivariant operation, and image position does not affect its result. A recent study shows that the zero-padding employed in convolutional layers of CNNs provides position information to the CNNs. The study further claims that the position information enables accurate inference for several tasks, such as object recognition, segmentation, etc. However, there is a technical issue with the design of the experiments of the study, and thus the correctness of the claim is yet to be verified. Moreover, the absolute image position may not be essential for the segmentation of natural images, in which target objects will appear at any image position. In this study, we investigate how positional information is and can be utilized for segmentation tasks. Toward this end, we consider {\em positional encoding} (PE) that adds channels embedding image position to the input images and compare PE with several padding methods. Considering the above nature of natural images, we choose medical image segmentation tasks, in which the absolute position appears to be relatively important, as the same organs (of different patients) are captured in similar sizes and positions. We draw a mixed conclusion from the experimental results; the positional encoding certainly works in some cases, but the absolute image position may not be so important for segmentation tasks as we think.
\end{abstract}

\section{Introduction}

Convolution is an equivariant operation, so are CNNs consisting only of convolutional layers \cite{lenc2019}. When shifting an input image spatially, the output will be shifted accordingly. In this ideal case, the absolute image position does not affect the result of the convolution. On the other hand, it could potentially be useful for solving various tasks including object recognition and segmentation. It has, however, not been well understood whether or not CNNs use positional information for these tasks, and if they do, how they use it.

Recently, an answer to this question was given by Islam et al. \cite{islam2020much}. They found that position information is provided to the convolutional layers by their (zero-)padding, and thus CNN may use it for standard tasks such as object recognition. Considering that padding is originally intended to adjust the size of the output maps, it is interesting to see that it unexpectedly plays an important role. 

In \cite{islam2020much}, the authors only analyze {\em existing} CNNs trained for standard tasks to see if and how they utilize position information. Then, a natural question arises: if position information is so important, {\em are there better ways to utilize it?} While zero padding provides position information only {\em implicitly}, we can provide it more explicitly by embedding it into the inputs. This idea is rather standard in the field of natural language processing,  
which is called {\em positional encoding}; when inputting a sequence of words, the relative position of each word is encoded and added to the word's feature. In this study, we consider a similar method tailored for image tasks that adds channels to the input image that encodes image coordinates. As far as the authors know, there is no study, despite its simplicity, that conducts a systematic analysis on the effectiveness of this method. There are a few studies that employ it \cite{DBLP:conf/miccai/TilborghsDCBM19}, but we are not aware of those providing detailed analysis of its effects. A few other studies attempt to utilize position information in a more sophisticated way, such as the use of recurrent networks \cite{visin2015renet} and capsule \cite{sabour2017dynamic}. Another study \cite{larrazabal2019anatomical} proposes to use a denoising auto-encoder to memorize the shape of segmentation masks as `anatomical prior' of human lungs and hearts. These studies do not provide a direct answer to the above question.

The study of Islam et al. \cite{islam2020much} concludes that 
a standard CNN (i.e., VGG-16) utilizes absolute position information for object recognition and segmentation tasks. This is debatable, however. For one thing, there is a technical issue with their experiments. It is that they compared CNNs with and without padding to assess the impact of position information provided by the padding. This is problematic, since removing padding from CNNs will make the size of high-layer feature maps significantly smaller, harming its performance; its impact could be much larger than position information. 

For another, we need to be careful to judge how important position information is for solving a given task. In the case of natural images,
the importance of position information will have to largely depend on how they are captured. For example, if the camera can point at all directions in space,
where an object(s) appear in the image will have to be completely random. In this case, the absolute position should not be utilized for inference.
Thus, even when position information contributes to inference conducted for some datasets (e.g., ImageNet and PASCAL VOC), it may be fair to say that it is because of the bias existing in these datasets. In fact, as is pointed out in several studies \cite{conf/iclr/GeirhosRMBWB19,DBLP:conf/iclr/HendrycksD19}, the images contained in ImageNet have strong biases, for example, a target object(s) will appear at near the center with a large size---not surprisingly for snapshots. Should we regard it as something to be avoided (i.e., spurious statistics of training data causing domain shift), or as ``prior knowledge'' that will improve inference accuracy? 

In this study, we investigate how positional information is and can be utilized for segmentation tasks. Considering the above nature of natural images, we choose medical image segmentation tasks, in which the absolute position appears to be relatively important, as the same organs (of different patients) are captured in similar sizes and positions. As mentioned above, we consider {\em positional encoding} that adds channels embedding image position to the input images. We conduct a series of experiments to compare it with several padding methods in terms of inference accuracy. We evaluate segmentation methods not only on the standard test samples contained in the datasets we use but also on their synthetically degraded versions to simulate domain shift that we frequently encounter in real-world problems.



\section{Methods}

\subsection{Positional Encoding}

We consider the following method of embedding absolute image positions into additional channels of inputs to CNNs. The input image in our experiments is always gray-scale and thus has originally a single channel. We add two channels to it, which encode $x$ and $y$ coordinates with their pixel values, respectively. To be specific, for an image of size $W\times H$, we augment it to have three channels $[I_0(i,j),I_1(i,j),I_2(i,j)]$ at pixel $(i,j)$, where $i=0,\ldots,W-1$ and $j=0,\ldots,H-1$; $I_0$ stores the original pixel intensity, and $I_1$ and $I_2$ encode $x$ and $y$ coordinates as follows:
\begin{subequations}
\begin{align}
    I_1(i,j) &= \lambda \cdot i/(W-1), \\
    I_2(i,j) &= \lambda \cdot j/(H-1).
\end{align}
\end{subequations}
Thus, these channels store values from $0$ to $\lambda$, where $\lambda$ is a control parameter. We can use $\lambda$ to change the relative importance of the position information to pixel intensities, as will be shown later in our experiments.

\subsection{Datasets}

We select the two datasets of medical image segmentation that matches the purpose of this study, JSRT \cite{shiraishi2000development} and ACDC \cite{bernard2018deep}. These datasets are widely used in many studies \cite{isensee2017automatic, bernard2018deep, zotti2018convolutional, painchaud2019cardiac, novikov2018fully, frid2018improving, bonheur2019matwo,larrazabal2019anatomical}. 


\bigskip
\noindent
{\bf JSRT}~ This dataset was created and distributed by Japanese Society of Radiological Technology (JSRT) \cite{shiraishi2000development}. It contains 247 chest X-ray images with segmentation masks of lungs and hearts. There are four segmentation masks for each image, i.e., the background, right lung, left lung, and heart. The original size of the images is $2,048\times2,048$ pixels (0.175mm/pixel) and their brightness resolution is 12 bit. We resize them into $256\times 256$ pixels for the purpose of computational efficiency. This should not have any effect on our analysis. We split the dataset into 156 images for training, 23 for validation, and 68 for testing. The image brightness is normalized so that its distribution over all the training samples will have mean 0 and variance 1. The same normalization is applied to other images as well.

\bigskip
\noindent
{\bf ACDC}~ This dataset was created for Automated Cardiac Diagnosis Challenge (ACDC) \cite{bernard2018deep}. It contains 3D cine MRIs with segmentation masks of hearts of 100 subjects. There are four classes in the masks, i.e., the background, left and right ventricular cabities, and myocardium. We resize each slice image into $256\times 256$ pixels. We normalize the brightness of each slice image so that it will have mean 0 and variance 1. We split 100 subjects into 70 for training, 10 for validation, and 20 for testing. 

\subsection{Synthetic Image Degradation}

When training and test data are collected from different distributions, machine learning models trained on the former will attain only suboptimal inference performance on the latter. This problem, also known as domain shift, has recently been recognized as a critical problem especially with deep learning.
The performance under such domain shift cannot be properly evaluated by splitting data collected by a single method into training and test splits. A solution that is effective for computer vision tasks is to synthetically deteriorate the quality of input images. By degrading the quality of only images for testing, we can simulate some types of domain shift. We can synthesize various types of image degradation in a realistic fashion. This is one of a few feasible methods that can evaluate the real-world performance of deep learning models. 

In this study, we apply image degradation to the test splits of the above datasets. Following the study \cite{DBLP:conf/iclr/HendrycksD19} that examines the robustness of CNNs to image degradation on the standard object recognition task, we choose the most realistic three types of degradation, i.e., {\em Gaussian noise}, {\em Gaussian blur}, and {\em shot noise},  out of thirteen proposed in the study. The parameter of the Gaussian noise is set as $\sigma_n=0.04$, 0.08, 0.12, 0.15, and 0.18 for normalized pixel intensity in the range $[0,1]$; the parameter of the Gaussian blur is set as $\sigma_n=0.04$, 0.08, 0.12, 0.15, and 0.18 pixels; and that for the shot noise is set as $c=15$, 30, 50, 100, and 250. 


\begin{figure}\small
\centering
\begin{tabular}{cc} 
\bmvaHangBox{\includegraphics[width=1.93cm]{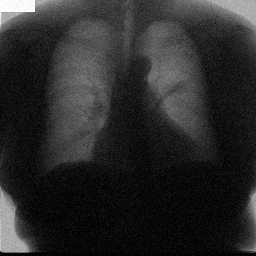}}
\bmvaHangBox{\includegraphics[width=1.85cm]{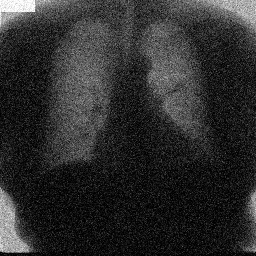}}
\bmvaHangBox{\includegraphics[width=1.85cm]{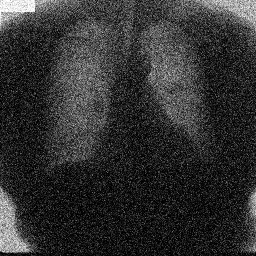}}&
\bmvaHangBox{\includegraphics[width=1.85cm]{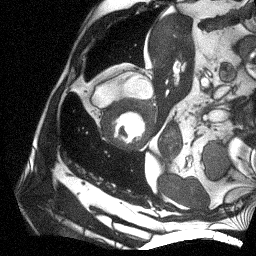}}
\bmvaHangBox{\includegraphics[width=1.85cm]{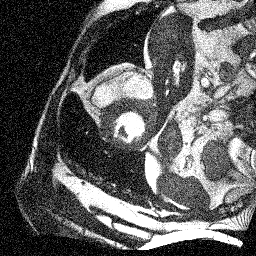}}
\bmvaHangBox{\includegraphics[width=1.85cm]{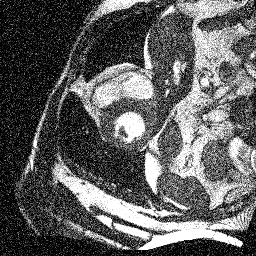}}\\
\bmvaHangBox{\includegraphics[width=1.85cm]{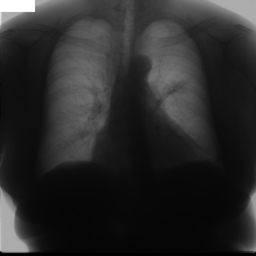}}
\bmvaHangBox{\includegraphics[width=1.85cm]{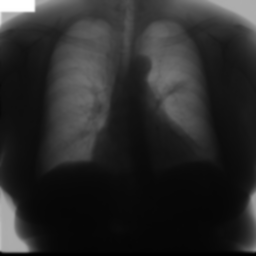}}
\bmvaHangBox{\includegraphics[width=1.85cm]{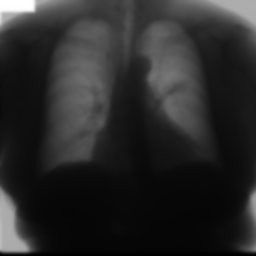}}&
\bmvaHangBox{\includegraphics[width=1.85cm]{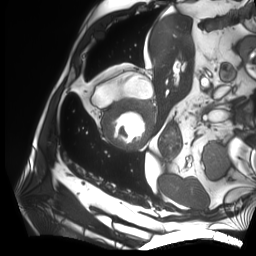}}
\bmvaHangBox{\includegraphics[width=1.85cm]{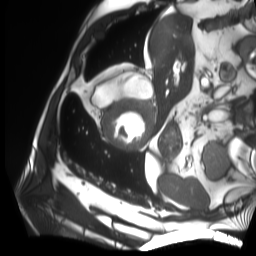}}
\bmvaHangBox{\includegraphics[width=1.85cm]{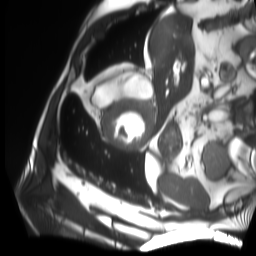}}\\
\bmvaHangBox{\includegraphics[width=1.85cm]{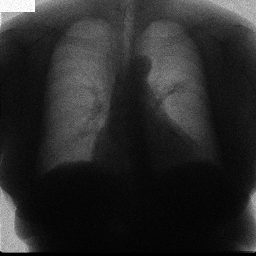}}
\bmvaHangBox{\includegraphics[width=1.85cm]{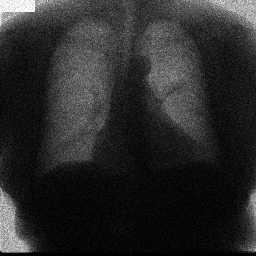}}
\bmvaHangBox{\includegraphics[width=1.85cm]{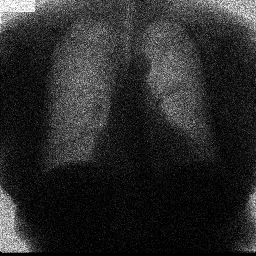}}&
\bmvaHangBox{\includegraphics[width=1.85cm]{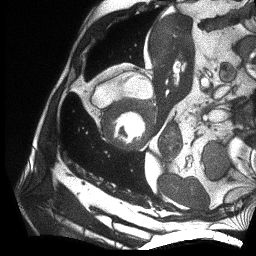}}
\bmvaHangBox{\includegraphics[width=1.85cm]{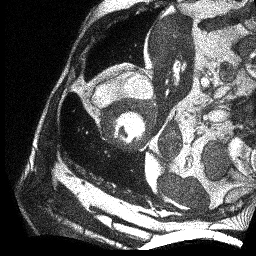}}
\bmvaHangBox{\includegraphics[width=1.85cm]{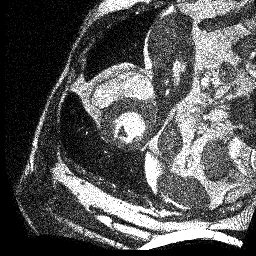}}\\
(a) JSRT & (b) ACDC
\end{tabular}
\caption{Example images from the two datasets and their degraded versions. From top to bottom rows, Gaussian noise, Gaussian blur, and shot noise. }
\label{fig:noise_example}
\end{figure}

\subsection{Experimental Setup}

For CNNs, we use a standard model, U-net \cite{ronneberger2015u} with batch normalization \cite{ioffe2015batch}, and a light-weigth CNN with only four conv. layers. We also use their variants having three input channels when used with the positional encoding. For their training and evaluation, we use standard methods used in the studies of medical image segmentation. To be specific, we train the CNNs using the standard cross-entropy loss and the Adam optimizer \cite{kingma2014adam} for 100 epochs. We set its learning rate to 0.001 and the batch size to 20. To evaluate their inference performance, we employ the Dice coefficient, which is commonly used in the studies using the above two datasets. We use Pytorch \cite{paszke2019pytorch} for all experiments.


\section{Experimental Results}

\subsection{Effectiveness of Positional Encoding in Standard Setting}
\label{sec:std}

\begin{figure}[t]\fontsize{7.5pt}{0pt}\selectfont
\centering
\begin{tabular}{ccc} 
\bmvaHangBox{\includegraphics[width=5cm]{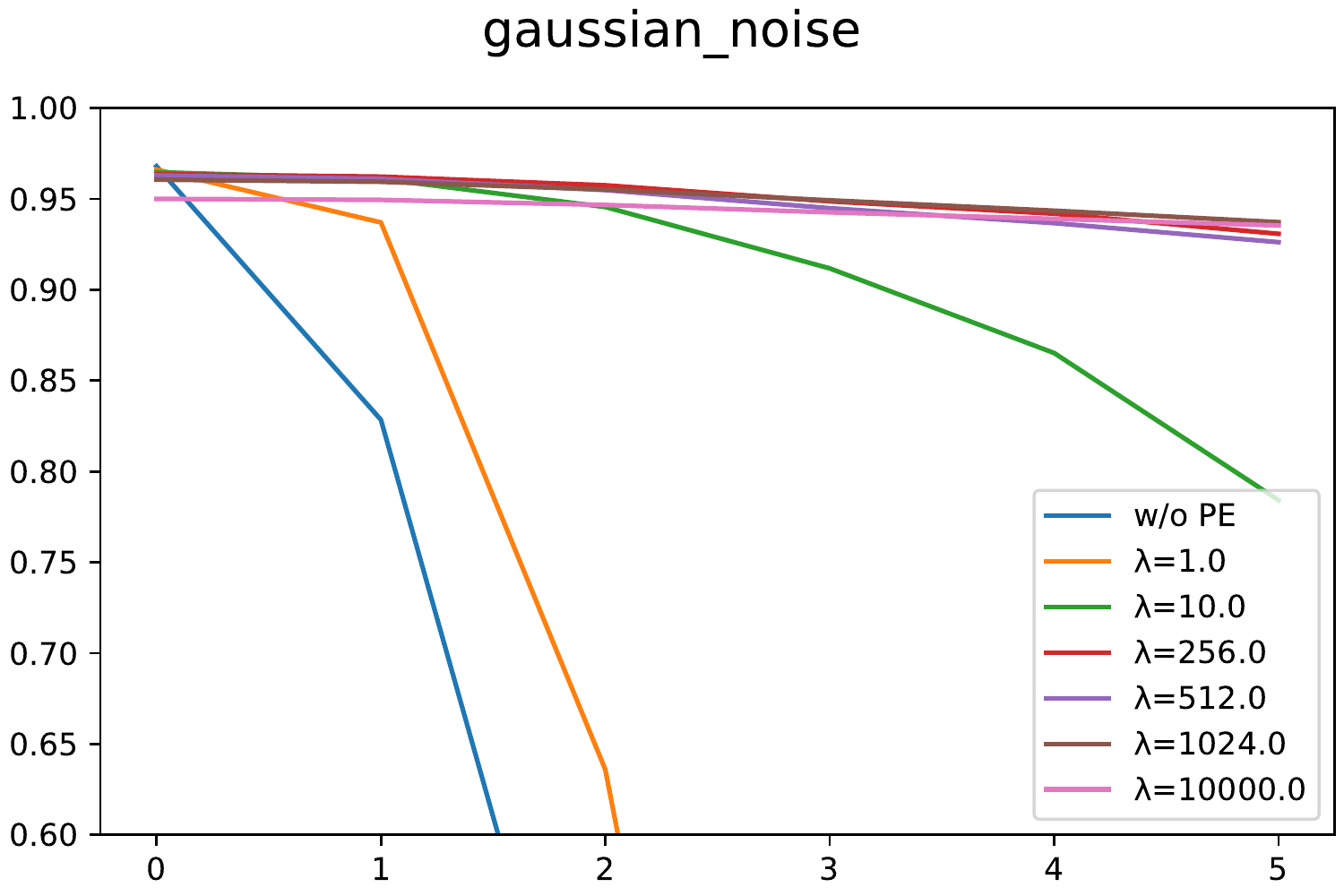}}&
\bmvaHangBox{\includegraphics[width=5cm]{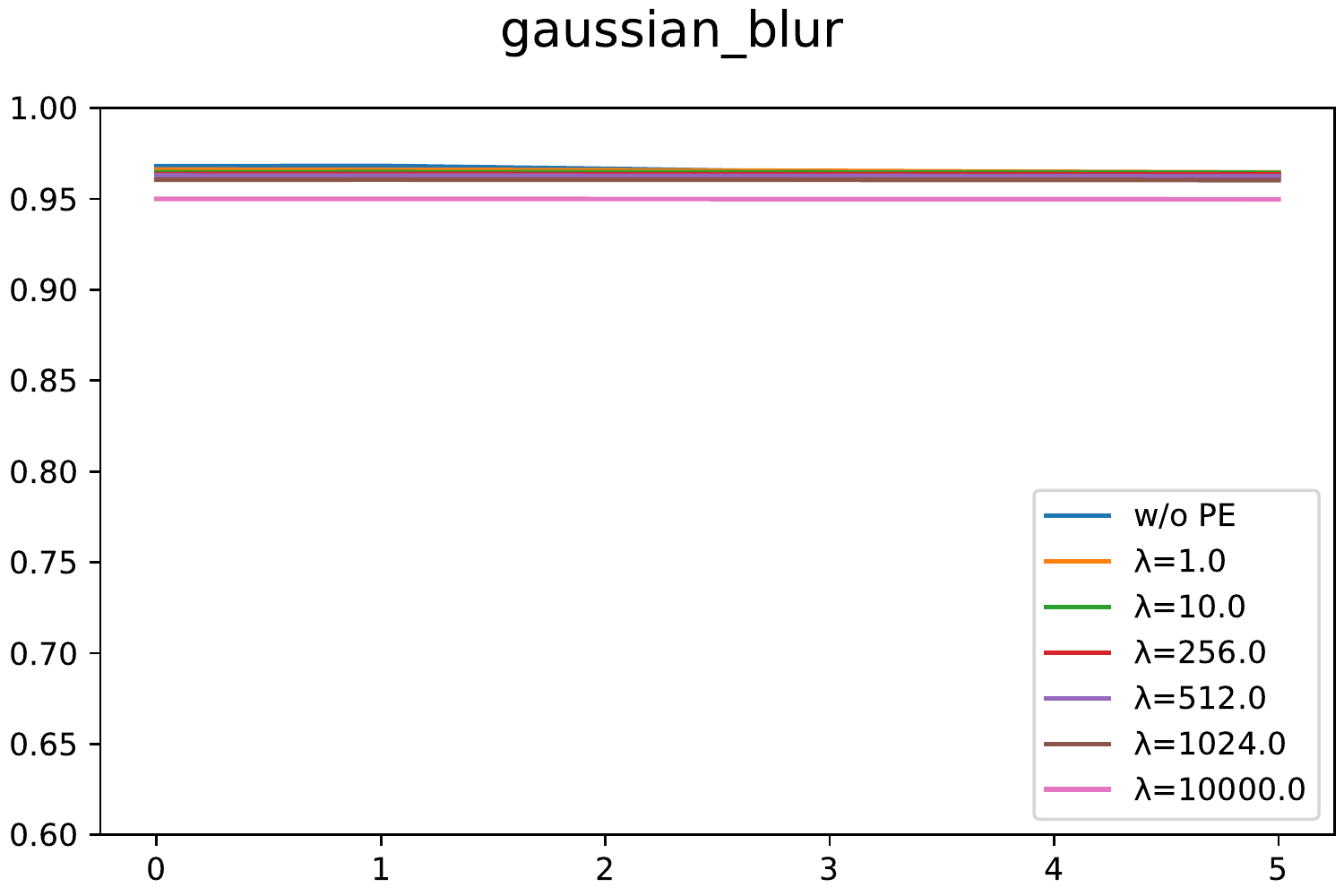}}&
\bmvaHangBox{\includegraphics[width=5cm]{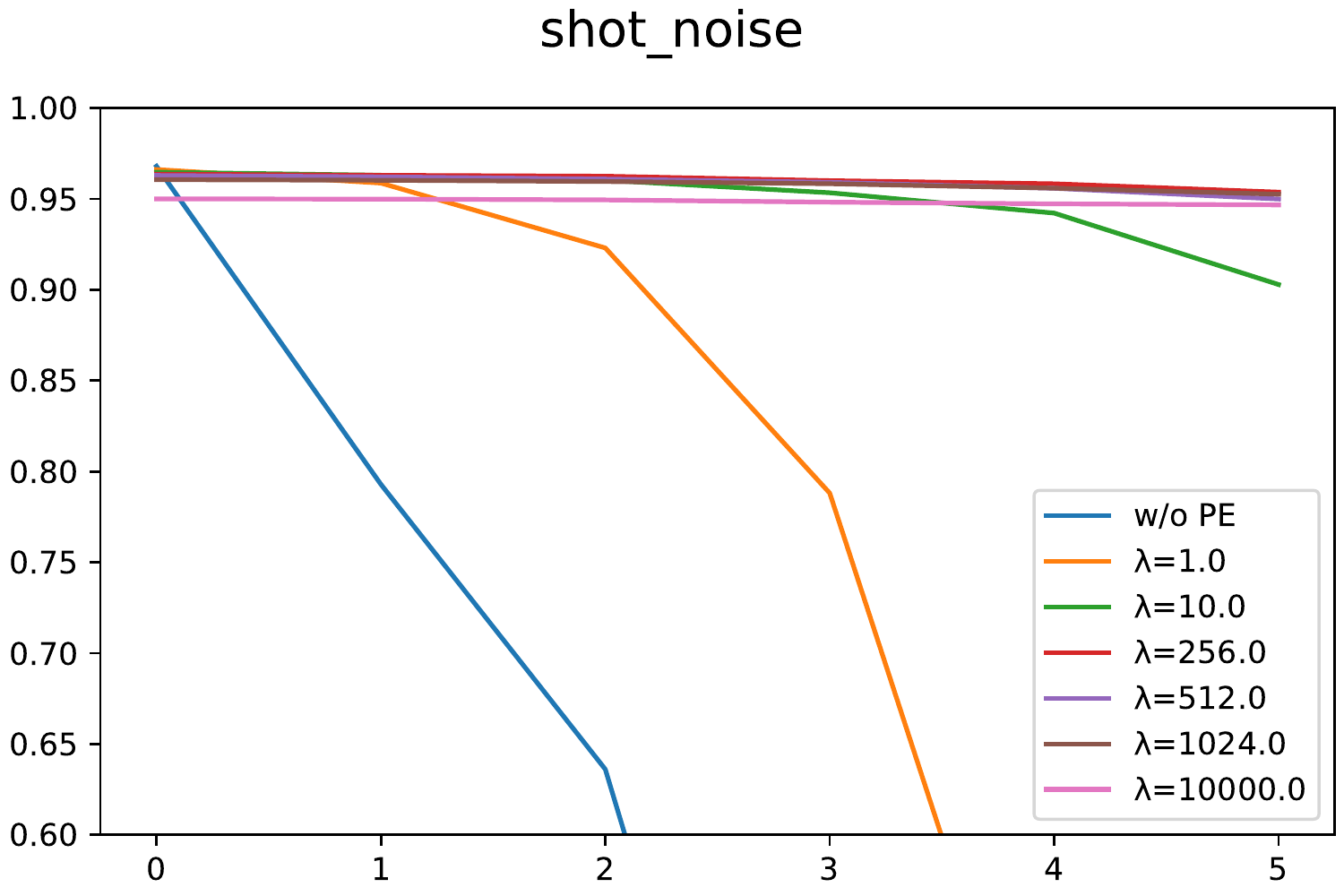}}\\
(a)&(b)&(c)
\end{tabular}

\vspace*{1mm}
\begin{tabular}{ccc} 
\bmvaHangBox{\includegraphics[width=5cm]{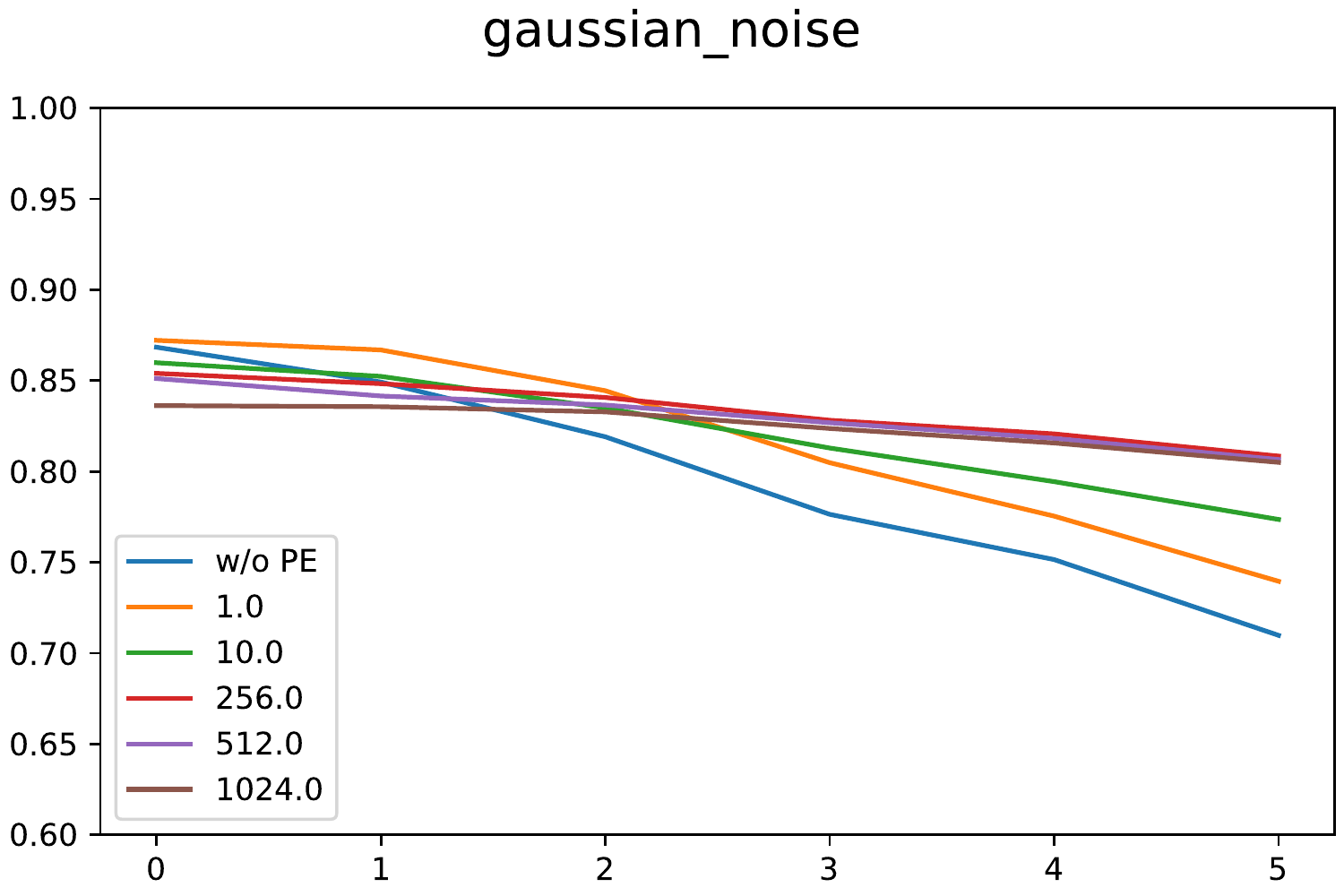}}&
\bmvaHangBox{\includegraphics[width=5cm]{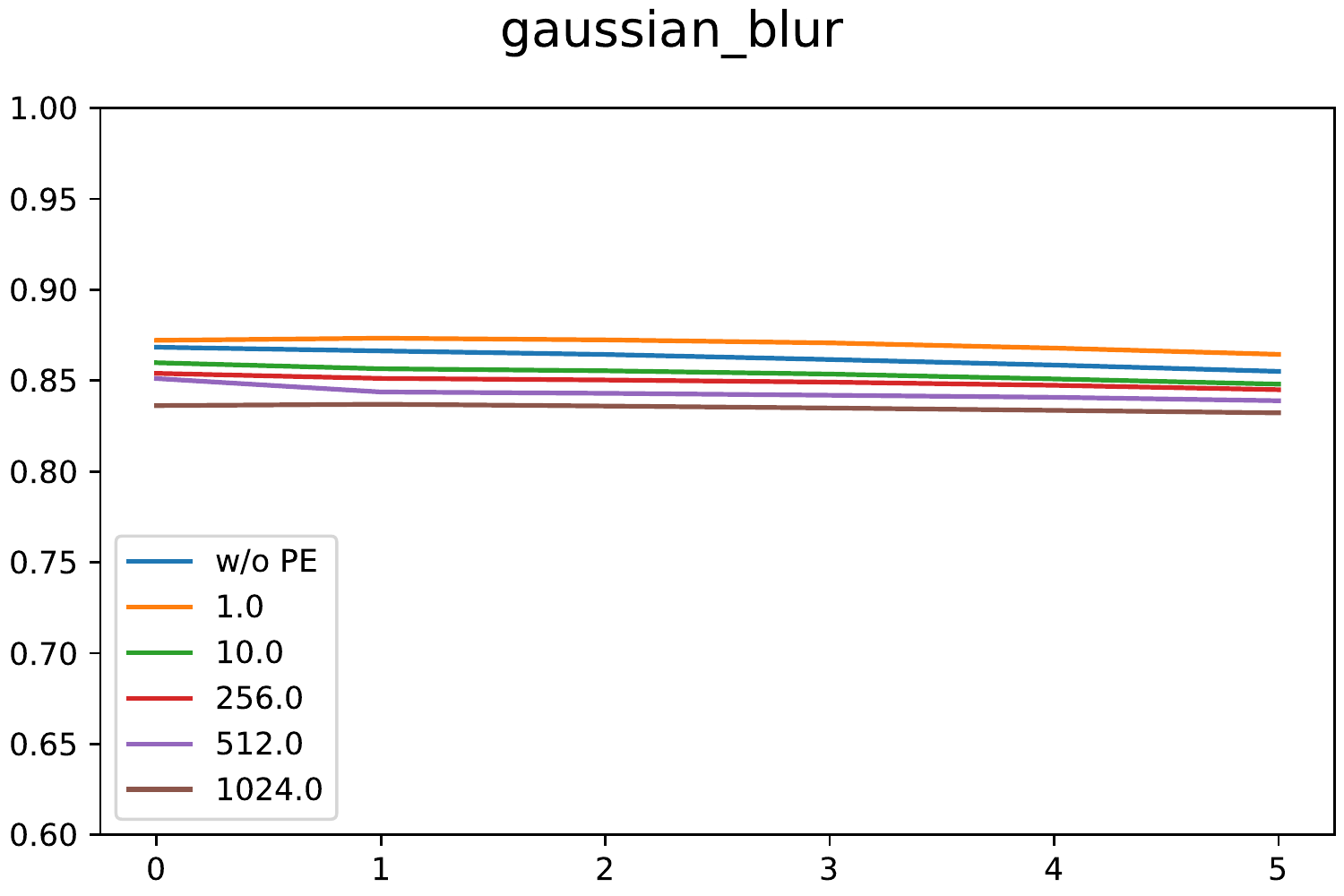}}&
\bmvaHangBox{\includegraphics[width=5cm]{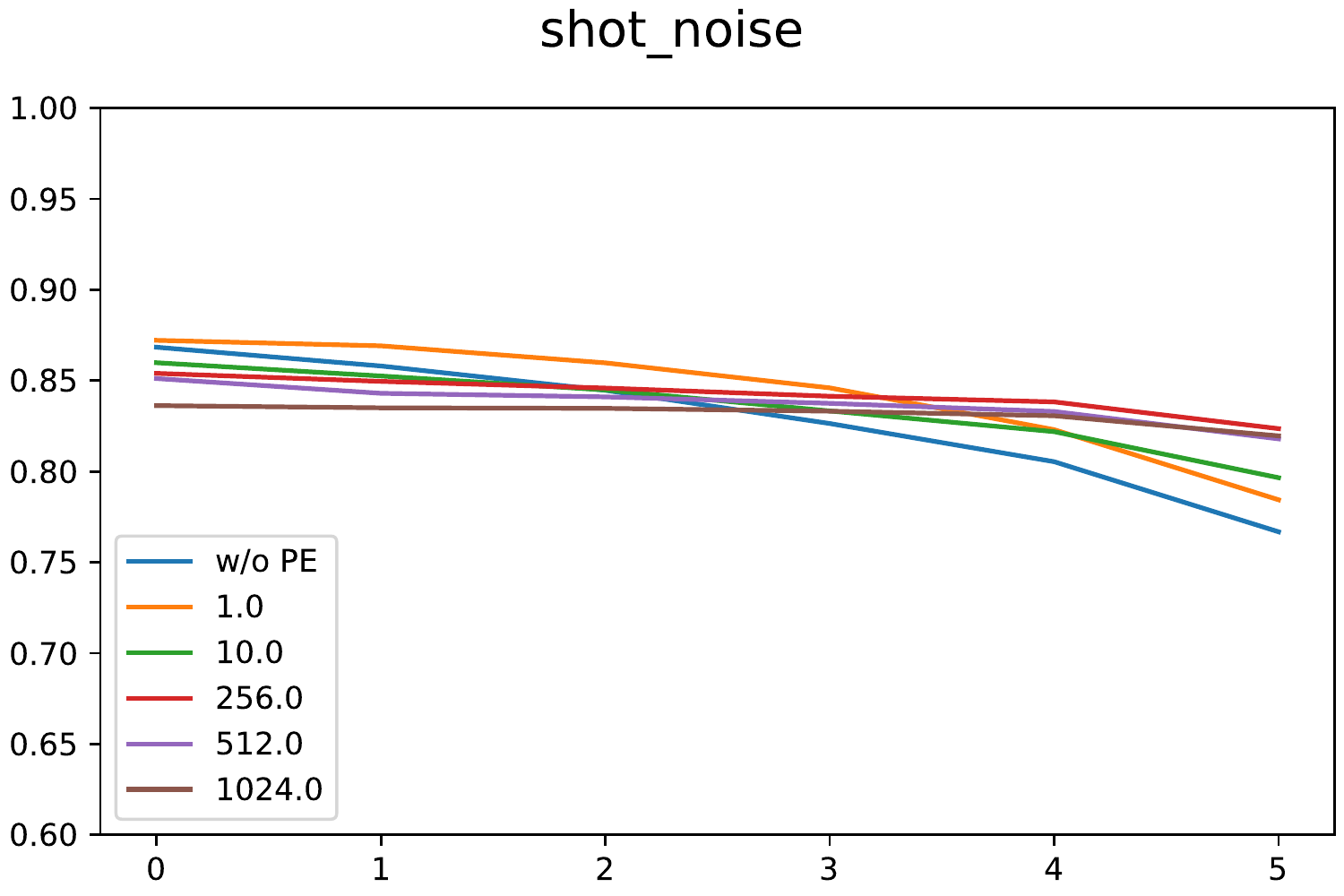}}\\
(d)&(e)&(f)
\end{tabular}

\vspace*{2mm}
\caption{Effects of positional encoding on the test splits of JSRT (upper row, (a)-(c)) and ACDC (lower row, (d)-(f)), to which three types of image degradation are applied. Positional encoding (PE) with different range $[0:\lambda]$ in $x$ and $y$ axes are employed at the input of U-net.}
\label{fig:jsrt_robustness}
\end{figure}

We first examine the effects of positional encoding (PE) in a standard setting. To be specific, we train a U-net model \cite{ronneberger2015u} with PE and that without PE on the full training split of each dataset explained above. We then test them on the test splits, to which we applied the aforementioned degradation effects. The details of the U-net model are given in the supplementary.
The model used here employs zero-padding at every convolutional layer. 

Figure \ref{fig:jsrt_robustness} shows the results. It is observed (curves in blue color) that the performance of the model without PE deteriorates quickly as degradation magnitude increases for two  degradation types except {\em Gaussian blur}. This holds true for the both datasets. These demonstrate {\em the vulnerability of the original model to domain shift caused by the degradation}. 
It is then seen that this performance deterioration is mitigated by the employment of PE. Thus, {\em PE does contribute to improvement in generalization over some types of unseen inputs}. 

It is also seen that the recovery from the performance deterioration caused by the image degradation vary greatly depending on the magnitude parameter $\lambda$ of PE. The recovery is only modest with small $\lambda$ and increases with $\lambda$. Further larger $\lambda$ improves accuracy for strong image degradation but sacrifices that for degradation-free inputs. This trade-off is seen in the results on the both datasets but more clearly seen for ACDC. The optimal value for $\lambda$ appears to depend on each dataset and degradation magnitude. 

Table \ref{tab:jsrt_clean} summarizes the results on degradation-free inputs. It is seen that the employment of PE yields rather slightly worse result for JSRT (this will be further investigated in Sec.~\ref{sec:jsrt_smalldata}) and a slight improvement for ACDC. In general, PE has only small impact on performance in the degradation-free setting.
We conjecture that in this setting,
the U-net model mostly achieves the upper-bound performance and there is no room for further improvements. 

\begin{table}[th]\footnotesize
\centering
\caption{
The performance for degradation-free inputs with different PE parameters $\lambda$'s. }
\smallskip
\begin{tabular}{cc}
{\small (a) JSRT} & {\small (b) ACDC} \\
\begin{tabular}{|c|c|}
\hline
Model & Dice  \\
\hline
Averaged Mask & 0.8124 \\
U-net w/o PE & ${\bf 0.9673 \pm 0.0011}$ \\
U-net w/ PE ($\lambda=1$) & $0.9649 \pm 0.0012$ \\
U-net w/ PE ($\lambda=10$) & $0.9637 \pm 0.0003$ \\
U-net w/ PE ($\lambda=256$) & $0.9599 \pm 0.0023$ \\
\hline
\end{tabular}&
\begin{tabular}{|c|c|}
\hline
Model & Dice  \\
\hline
Averaged Mask & 0.158 \\
U-net w/o PE & $0.8496 \pm 0.0133$ \\
U-net w/ PE ($\lambda=1.0$) & ${\bf 0.8582 \pm 0.0100}$ \\
U-net w/ PE ($\lambda=10.0$) & $0.8557 \pm 0.0035$ \\
U-net w/ PE ($\lambda=256.0$) & $0.8347 \pm 0.0161$ \\
\hline
\end{tabular}
\end{tabular}
\label{tab:jsrt_clean}
\end{table}

\subsection{Evaluation with a Small-size CNN Having Limited Capacity}
\label{sec:small}

\begin{figure}\fontsize{7.5pt}{0pt}\selectfont
\centering
\begin{tabular}{ccc}
\bmvaHangBox{\includegraphics[width=5cm]{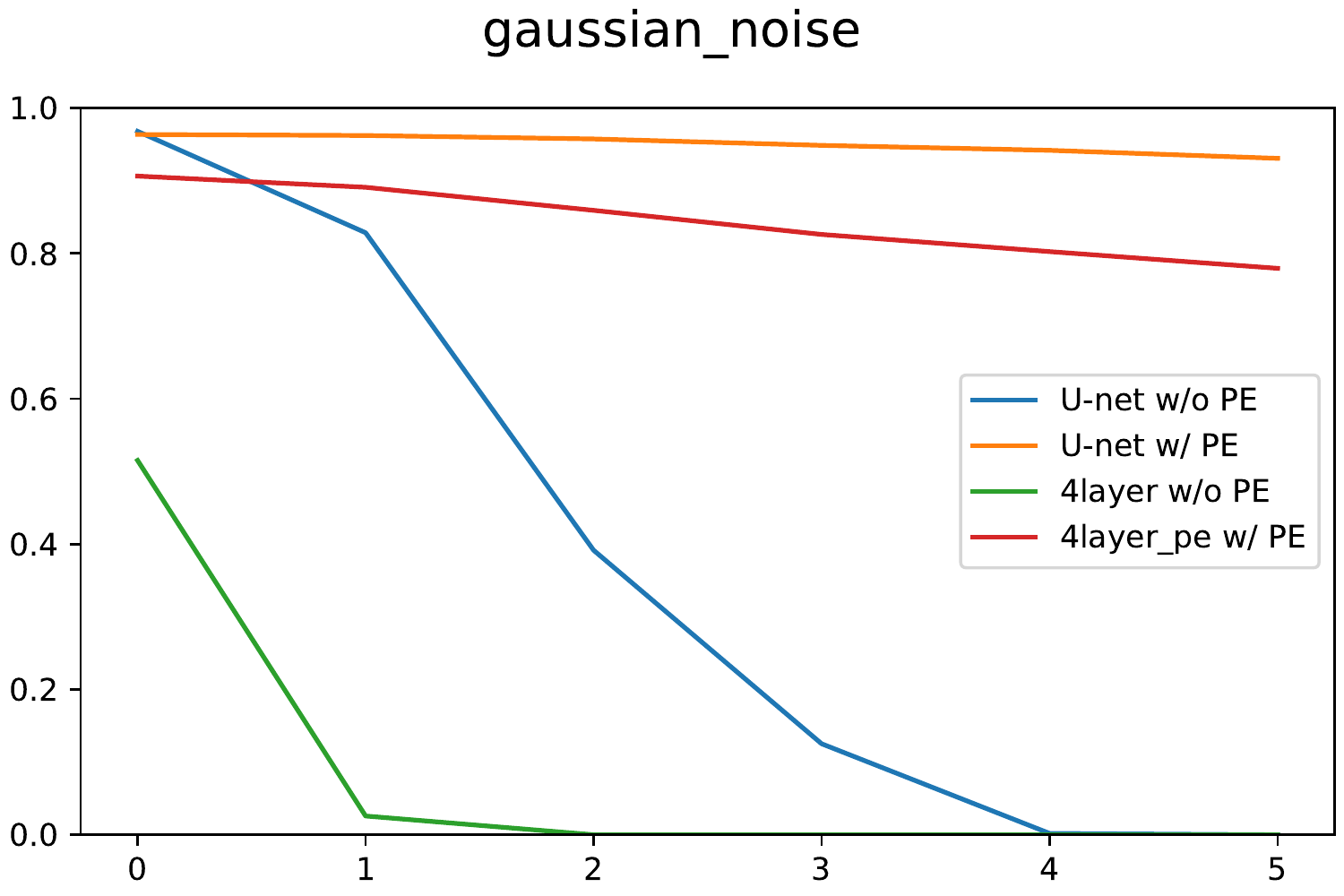}}&
\bmvaHangBox{\includegraphics[width=5cm]{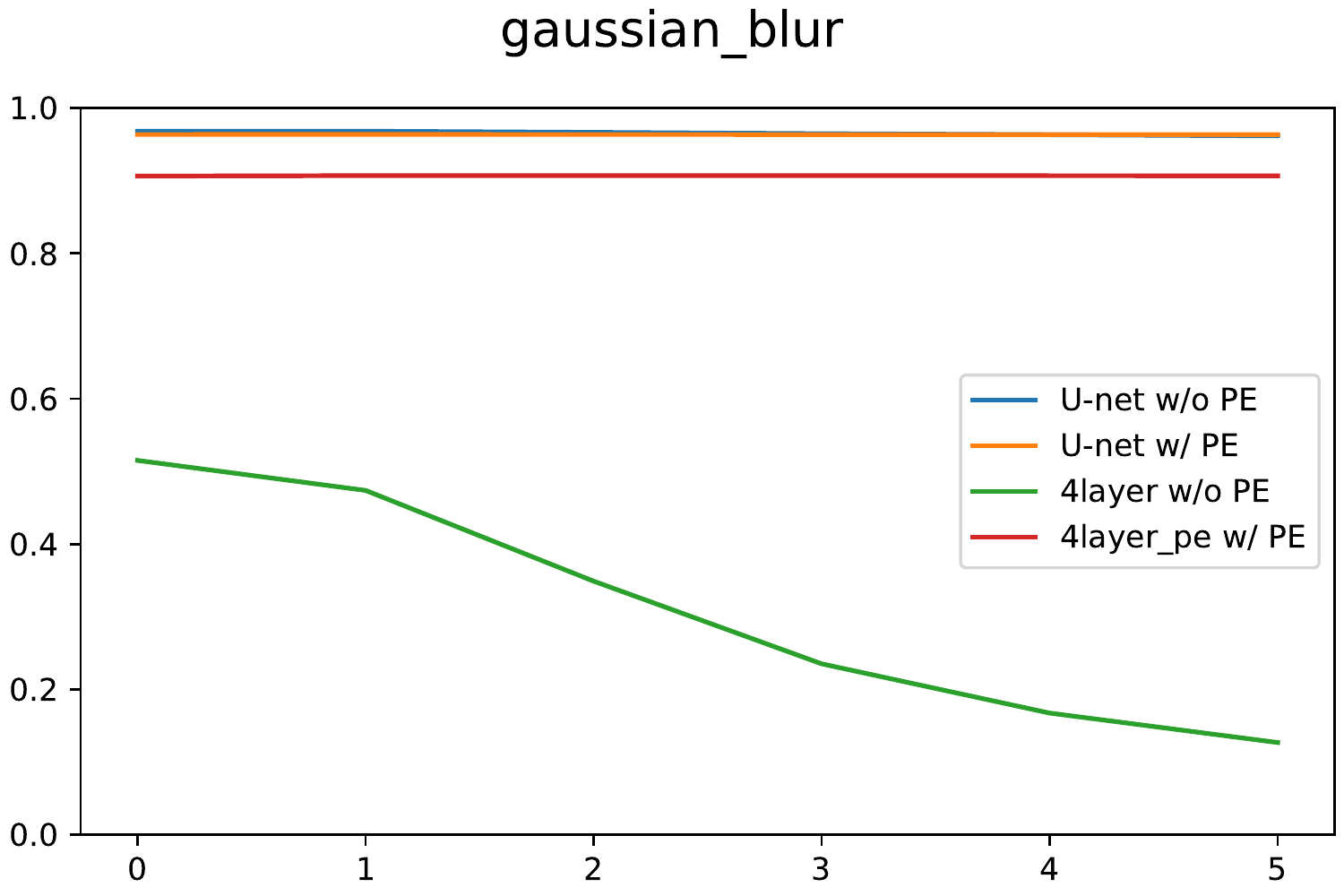}}&
\bmvaHangBox{\includegraphics[width=5cm]{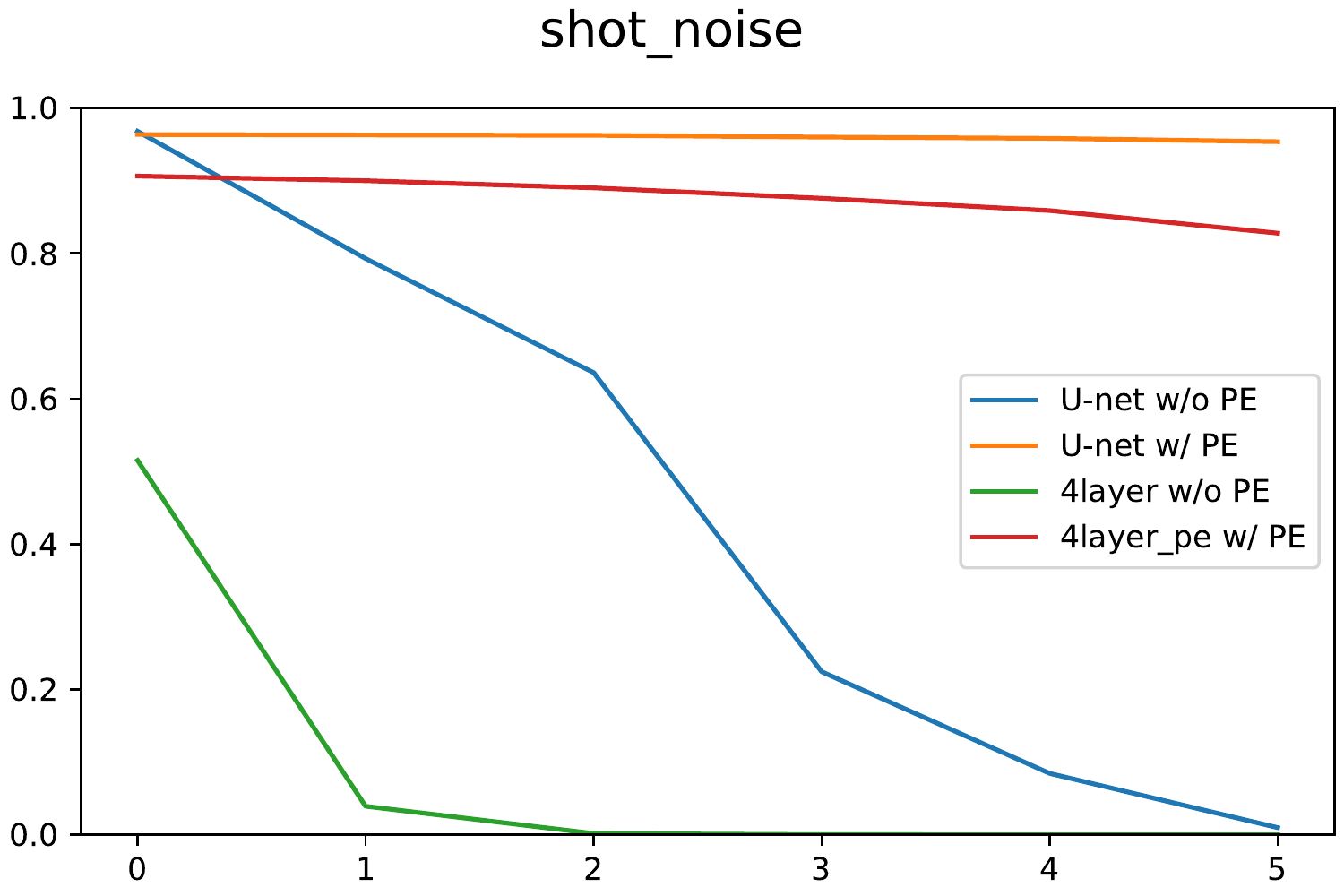}}\\
(a)&(b)&(c)
\end{tabular}

\vspace*{1mm}
\begin{tabular}{ccc}
\bmvaHangBox{\includegraphics[width=5cm]{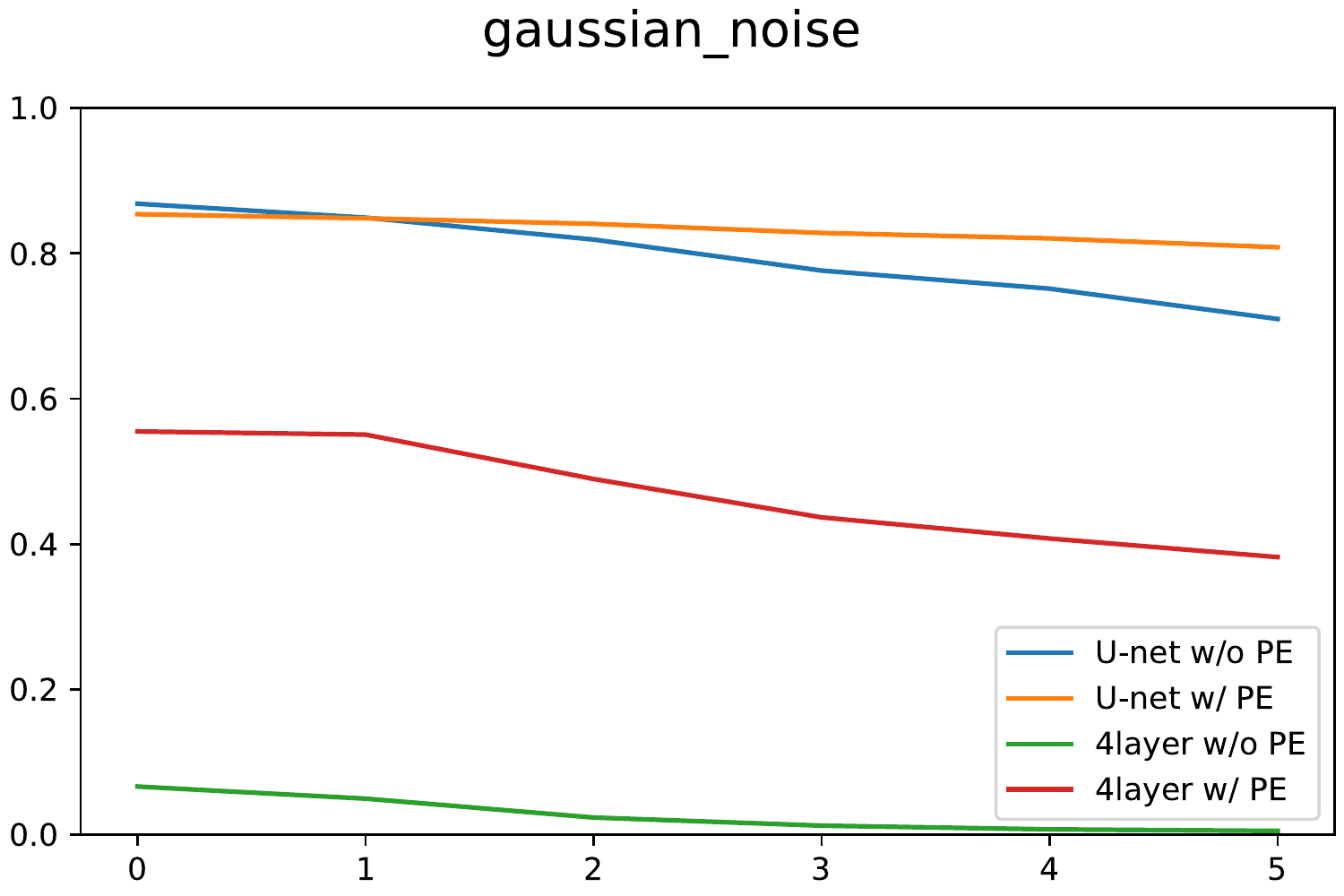}}&
\bmvaHangBox{\includegraphics[width=5cm]{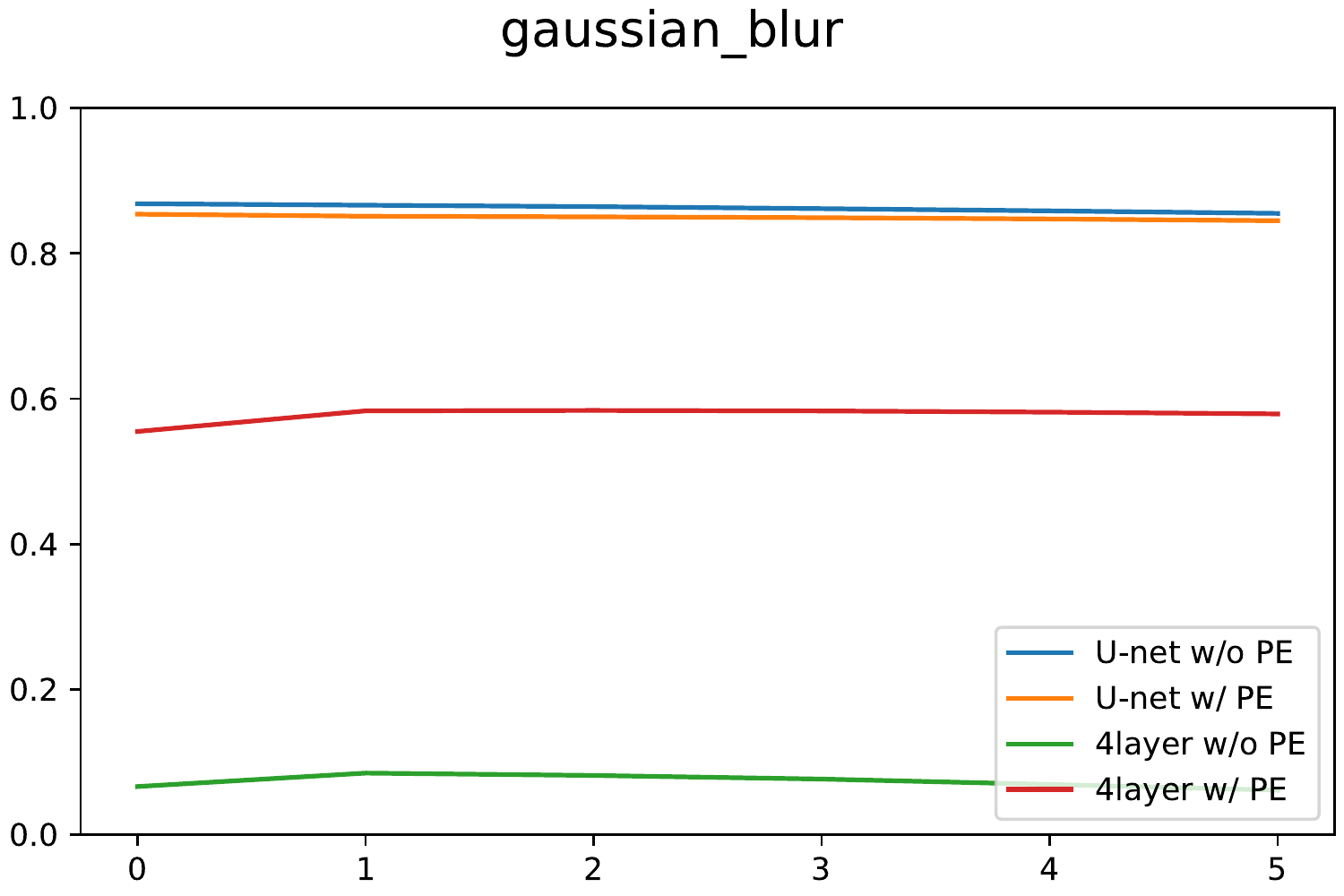}}&
\bmvaHangBox{\includegraphics[width=5cm]{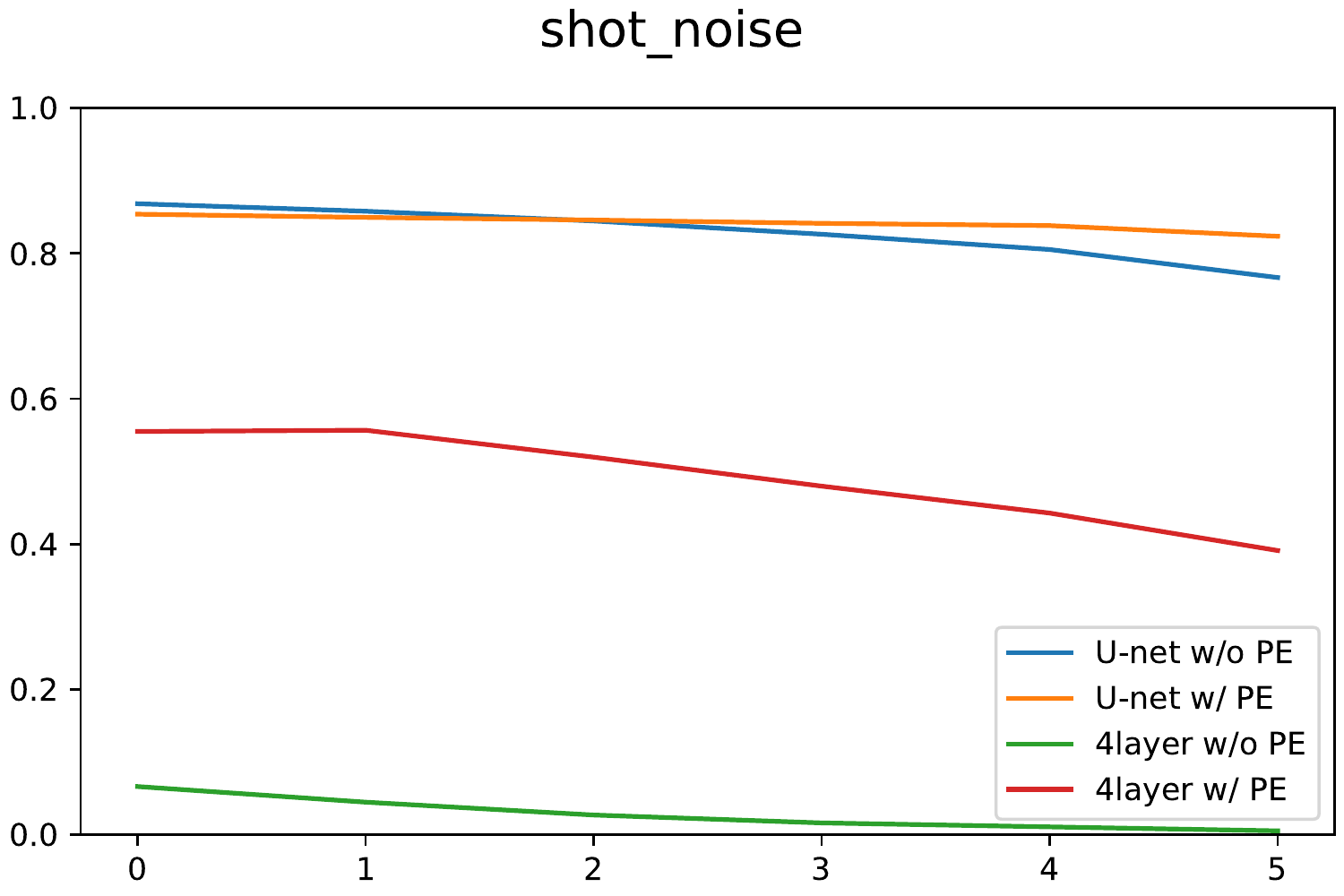}}\\
(d)&(e)&(f)
\end{tabular}

\vspace*{2mm}
\caption{Effects of positional encoding on a small-size CNN, i.e., a four layer CNN; see texts for details. Results on JSRT and ACDC are shown in the upper row (a)-(c) and lower row (d)-(f), respectively.  }
\label{fig:jsrt_small}
\end{figure}

To verify the above conjecture that the U-net model already achieves the upper-bound performance, we evaluate the performance of a smaller-size CNN that is intentionally given limited capacity. To be specific, we consider a four-layer CNN (its details are given in the supplementary material). We follow the same procedure as above, i.e., training it on clean images from the training split and testing on the same set of degraded images. The CNN employs the same zero-padding at every convolutional layer. We set $\lambda=256$ for PE.

Figure \ref{fig:jsrt_small} shows the results. It is seen that this small-size model, when used without PE, performs very badly even for clean images; accuracy is only slightly above 0.5 on JSRT and below 0.1 on ACDC, respectively. Then, the employment of PE significantly improves its performance, i.e., about 0.9 and 0.55 for JSRT and ACDC, respectively. Note that these are much better than a naive `baseline' of calculating the average of segmentation masks over the training samples and using it as prediction, which attains 0.8124 and 0.158 for JSRT and ACDC, respectively; these are reported in Table \ref{tab:jsrt_clean}. This indicates that this small-size CNN has successfully learned to utilize both position information and pixel intensity effectively for inference. PE increases robustness to image degradation as well, as seen in the results of U-net. We can conclude from these results that PE is effective not only in increasing robustness to image degradation but also in the degradation-free setting; it will be effective in the conservative scanario of training and testing on data from an identical distribution.


\subsection{Evaluation with Reduced Training Data}
\label{sec:jsrt_smalldata}


\begin{table}\footnotesize
\centering
\caption{
Relation between performance (Dice coefficient) for degradation-free inputs and the size of training data. U-net with or without PE is trained on JSRT. }
\smallskip
\begin{tabular}{|c|c|c|}
\hline
Training data & W/O PE & W/ PE  \\
\hline
156 images & ${\bf 0.9673 \pm 0.0011}$ & $0.9599 \pm 0.0023$\\
10 images & $0.8813 \pm 0.0175$ & ${\bf 0.9020 \pm 0.0027}$\\
5 images & $0.8700 \pm 0.0170$ & ${\bf 0.8774 \pm 0.0055}$\\
\hline
\end{tabular}
\label{tab:jsrt_smalldata}
\end{table}

As noted above, it is seen in Table \ref{tab:jsrt_clean} that the use of PE with the U-net model does not improves performance on JSRT in the degradation-free setting (i.e., clean images). We conjectured that this is because the U-net model without PE has already achieved the optimal performance. This is partly verified by the above results with the small-size model. We consider another scenario in which the U-net model cannot achieve an ideal performance, i.e., when the size of training data is insufficient; this could create room for improvement by PE. 

Thus, we evaluate the performance of the same U-net model trained on a reduced set of training data. To be specific, we reduce the number of training samples from 156 that are used in the above experiments to 10 or 5 by random sampling. Table \ref{tab:jsrt_smalldata} shows the results. 
It is observed that the model without PE attains lower performances with the reduced training data, and they are improved by the employment of PE. Thus, PE is effective in the case of insufficient training data as well.

\subsection{Different Padding Methods}
\label{sec:padding}

It is argued in \cite{islam2020much} that zero-padding provides position information to CNNs. As a proof for this argument, the study shows the experimental results that CNNs with padding perform better on object recognition and detection than those without padding; specifically, a VGG-16 model with and without padding is trained and tested on those tasks (i.e., ImageNet and PASCAL VOC). 

However, removing padding from all convolutional layers from a CNN such as VGG-16 will significantly reduce the spatial resolution of the output maps of higher layers. Specifically, that of {\tt pool5} of VGG-16 is originally $7\times 7$, whereas that of the modified, padding-free VGG-16 will be $1\times 1$. 
It is no wonder that such an architectural change leads to decreases in inference accuracy, and thus the results do not provide a proof for the argument. It still remains unclear how important position information is for CNNs to make accurate inference on a variety of tasks.

There are several methods of padding, such as zero-, wrap-, clamp- and reflection-padding. It has been a standard knowledge of image processing that different padding will have different effects on the image border; see Sec.~3.2 of \cite{szeliski2011computer} for instance. Thus, we examine here what will happen when zero-padding in CNNs is replaced with reflection-padding.
Zero-padding applied to an input map to a conv. layer should generate artificial edges at the border of the resulting map, unless its values at the image border are exact zero. (This is not true in most cases.) It is then conjectured that such edges will provide clue for position information. In the case of reflection-padding, such artifical edges will not emerge; although it could still provide position information, its impact should be smaller than  zero-padding.


\begin{figure}\fontsize{7.5pt}{0pt}\selectfont
\centering
\begin{tabular}{ccc}
\bmvaHangBox{\includegraphics[width=5cm]{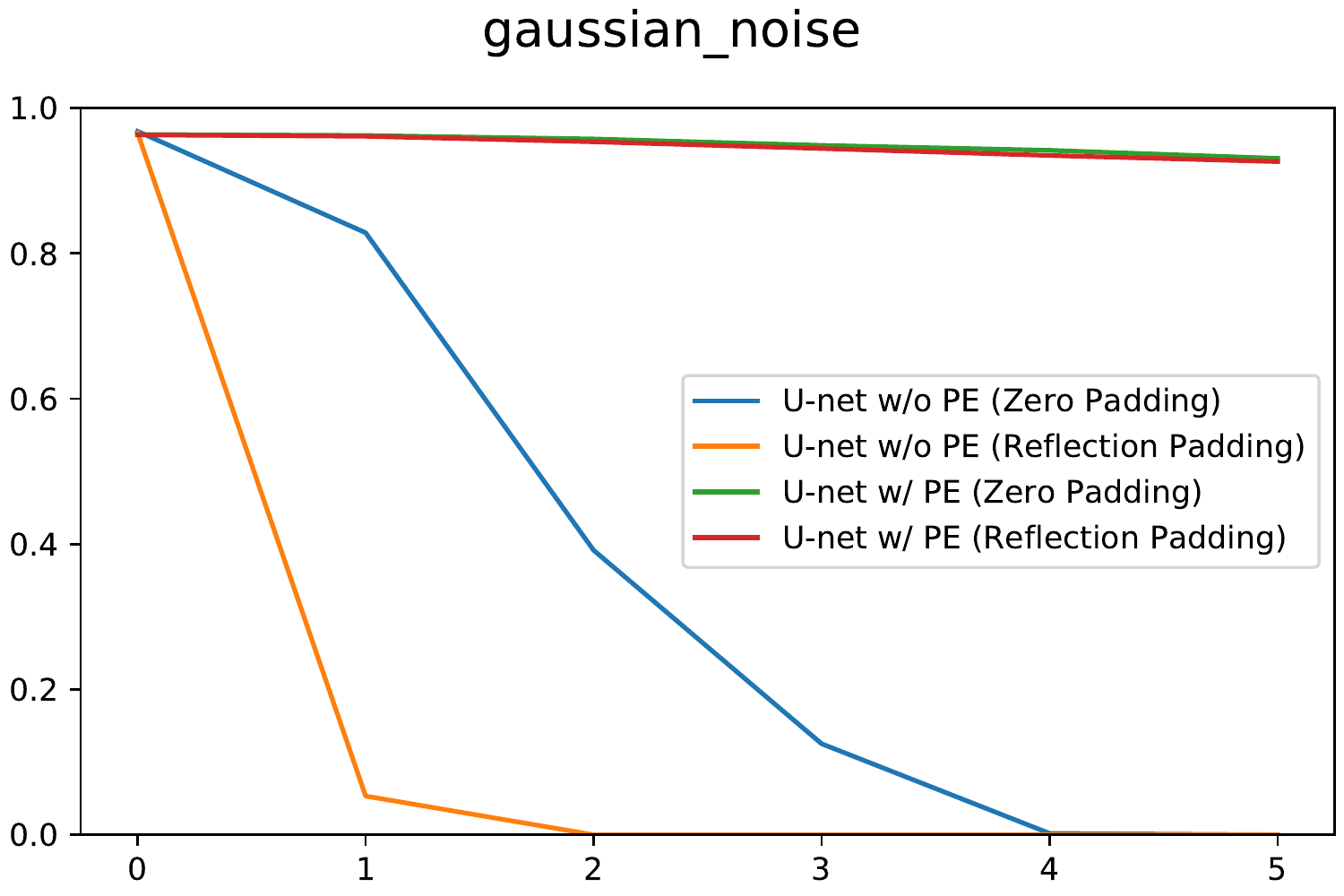}}&
\bmvaHangBox{\includegraphics[width=5cm]{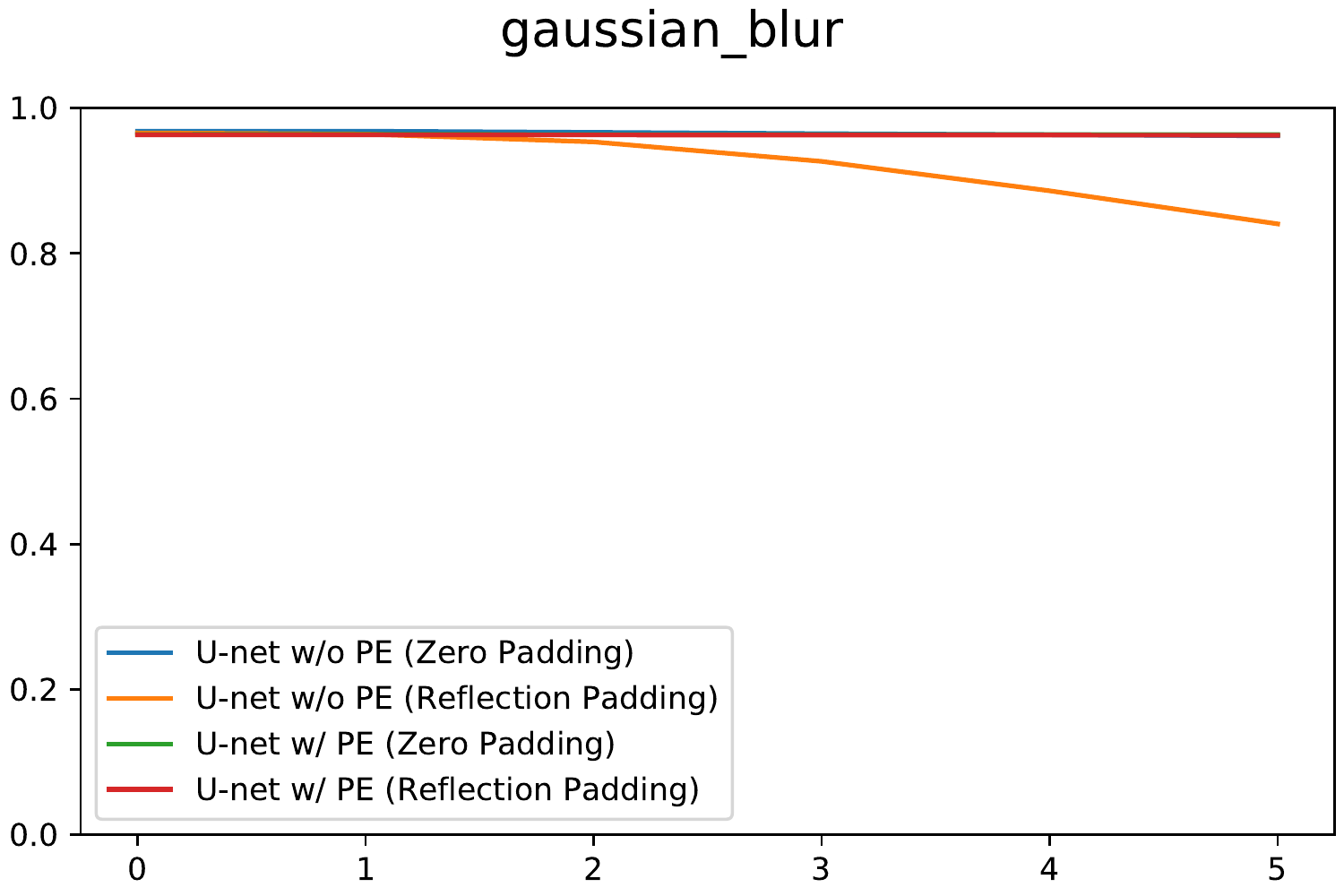}}&
\bmvaHangBox{\includegraphics[width=5cm]{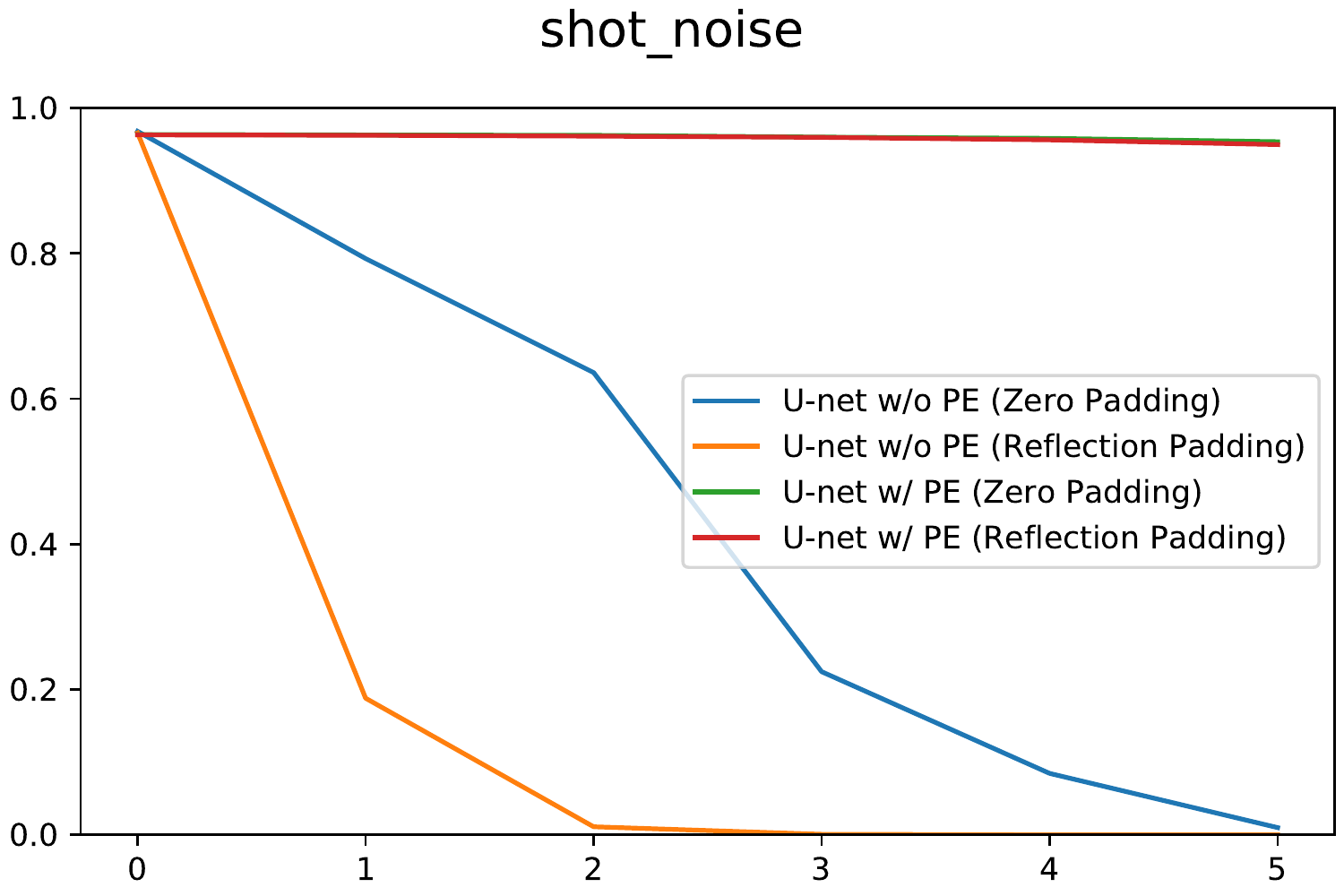}}\\
(a)&(b)&(c)
\end{tabular}

\vspace*{1mm}
\begin{tabular}{ccc}
\bmvaHangBox{\includegraphics[width=5cm]{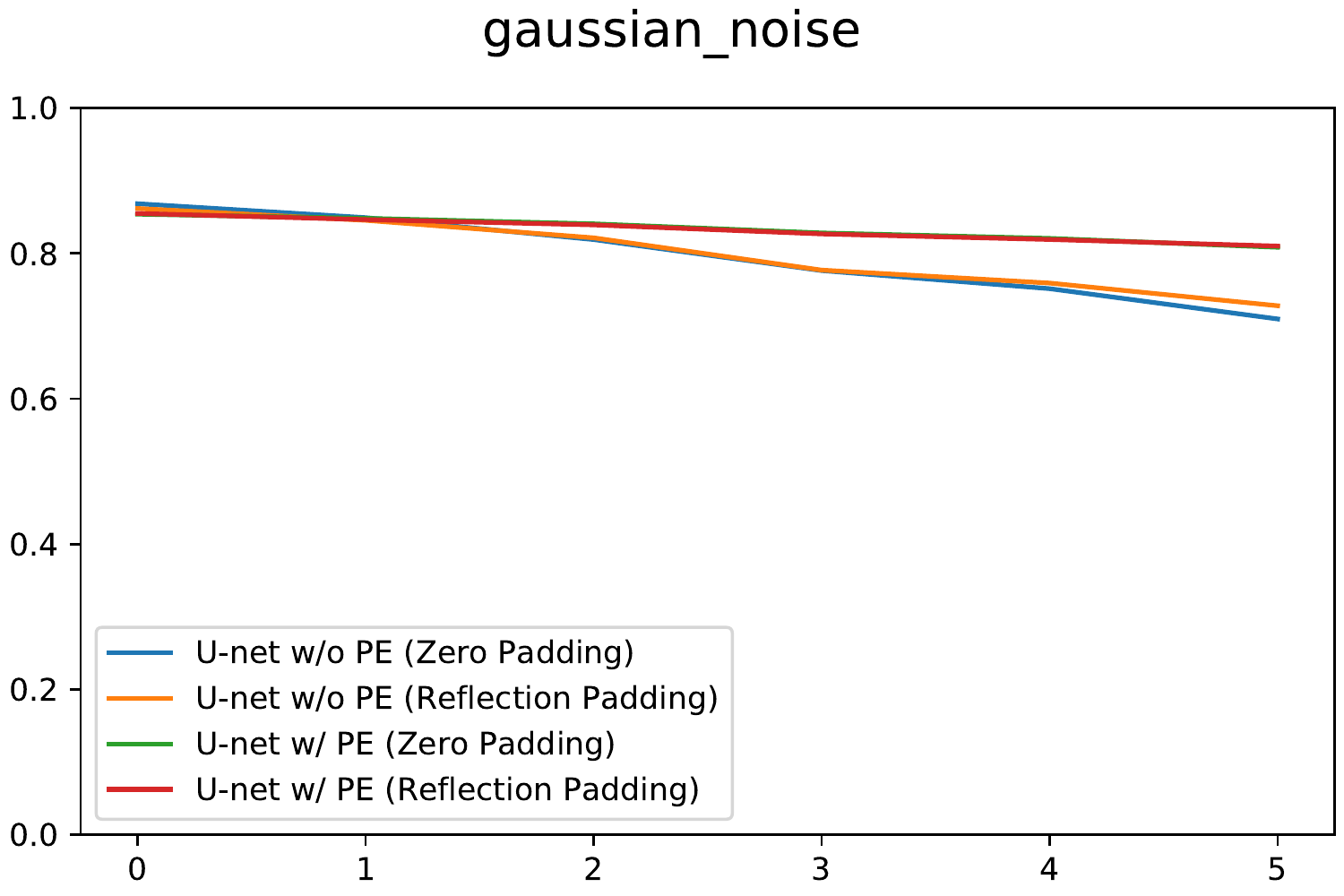}}&
\bmvaHangBox{\includegraphics[width=5cm]{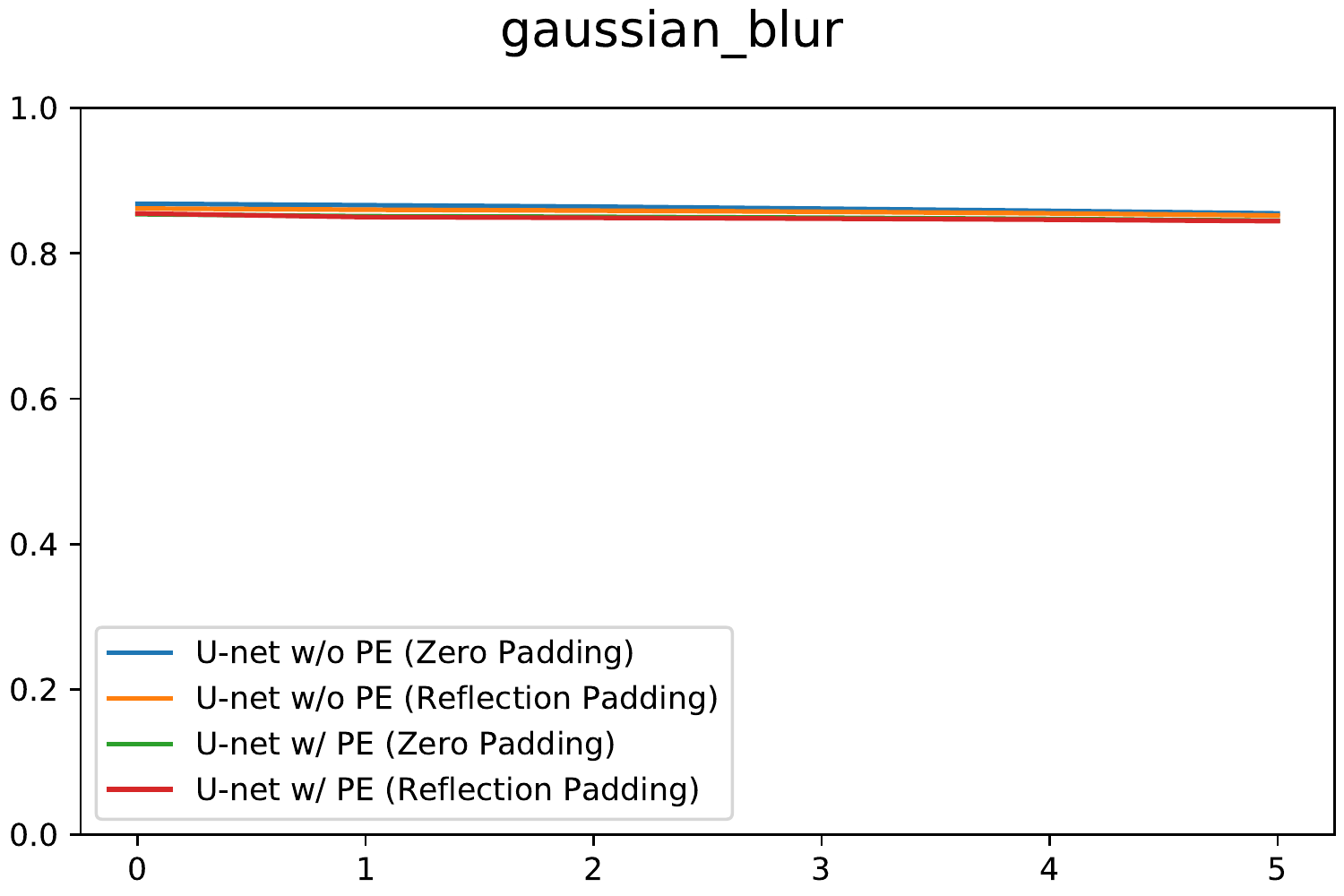}}&
\bmvaHangBox{\includegraphics[width=5cm]{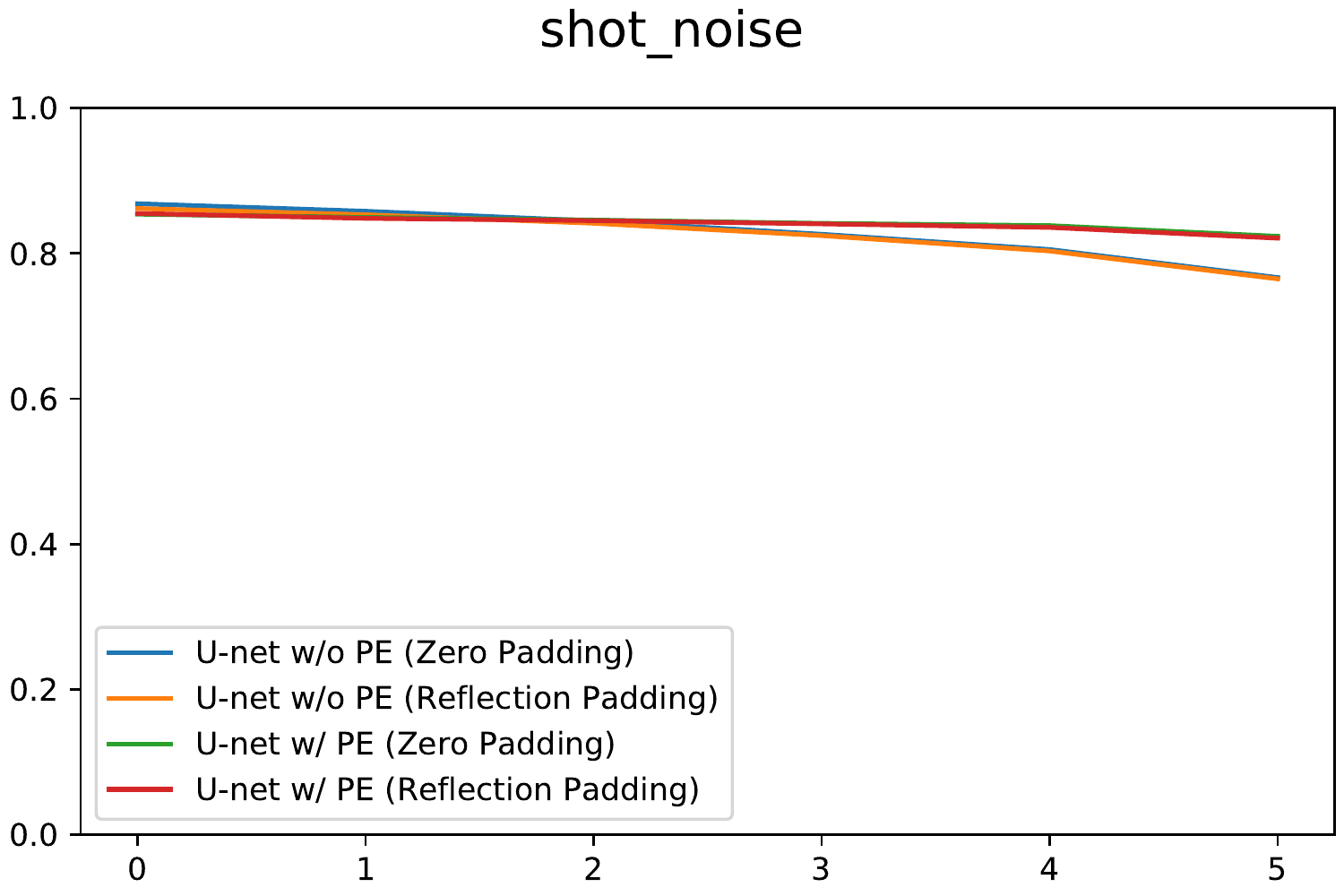}}\\
(c)&(d)&(e)
\end{tabular}

\vspace*{2mm}
\caption{Effects of positional encoding on different types of padding (i.e., zero padding and reflection padding). Upper row: JSRT. Lower row: ACDC. }
\label{fig:jsrt_padding}
\end{figure}

We repeat the same experimental procedure as above but with the same U-net model employing reflection-padding instead of zero-padding. Figure \ref{fig:jsrt_padding} shows the results. The difference in padding makes meaningful difference only for JSRT. In particular, when PE is not employed, the model with reflection-padding performs fairly worse than that with zero-padding in terms of robustness to degradation other than Gaussian blur. There is no difference in the performance of the degradation-free setting between the two padding methods. The incorporation of PE into the model with reflection-padding significantly improves the robustness to the same level attained by the model with zero-padding and PE. 

These results indicate that zero-padding does provide position information more effectively than reflection-padding, as is expected. Moreover, the difference in padding methods and with/without PE does not affect performance in the degradation-free setting. This implies that the U-net model does not rely much on position information for these datasets in the degradation-free setting. 

\subsection{Summary and Interpretation of the Results}

\noindent
{\bf PE improves robustness to degradation}~
{\em  
This is confirmed in the experiments of Sec.~\ref{sec:std}, \ref{sec:small}, and \ref{sec:padding}. }  The most likely explanation for this phenomenon is that the CNNs utilize the absolute position provided by PE to memorize the shape of the segmentation masks, leading to improved robustness to degradation, or equivalently, increased generalization over the induced domain shift. In fact, memorizing the average masks and using them for inference yield fairly accurate results on the both datasets, verifying the usefulness of this `prior knowledge'.  

\medskip
\noindent
{\bf Trade-off in degradation robustness with respect to $\lambda$}~
{\em PE with larger $\lambda$ attains more robustness to larger degradation, but it comes with a loss in accuracy for undegraded images.} Even the smallest $\lambda$ used in the experiment (i.e., $\lambda=1$) is enough for providing position information, as its magnitude is comparable to the range of the pixel intensities, i.e. $[-1:4]$ (after normalization); their distributions for the two datasets are given in the supplemnetary material. Why is PE with larger $\lambda$ nevertheless more effective for large input degradation? The most likely explanation is as follows. Its definition states that $\lambda$ controls the relative importance between the pixel intensity $I_0$ and absolute position $(I_1,I_2)$ contained in the augmented input signal $[I_0,I_1,I_2]$. Thus, it can control which of the two is utilized mainly for inference. A larger $\lambda$ will induce the CNN to learn to use position rather than pixel inetnsity. However, too much weight on position information will lead to performance decrease for undegraded images and those with a little degradation. 
 
\medskip
\noindent
{\bf Performance for undegraded images}~
{\em The employment of PE with the U-net model has practically no impact on the performance for undegraded images. For JSRT, it has even a slightly negative effect, but has a positive effect when training on a small amount of training data. Additionally, the use of PE with a small-size CNN having limited capacity has a positive effect on the both datasets.} These imply that the U-net model does not rely much on the absolute image position to make accurate inference at least for undegraded images. It should be noted that even if position information is available, it does not mean that CNNs learn to use it. As the CNNs are trained only on undegraded images in our experiments, they learn to simply ignore position information if it is unnecessary and instead utilize pixel intensity. %
\begin{wrapfigure}{r}[0pt]{0.43\textwidth}
\centering
\vspace*{-10pt} %
\includegraphics[clip,scale=0.37,bb=0 0 432 288]{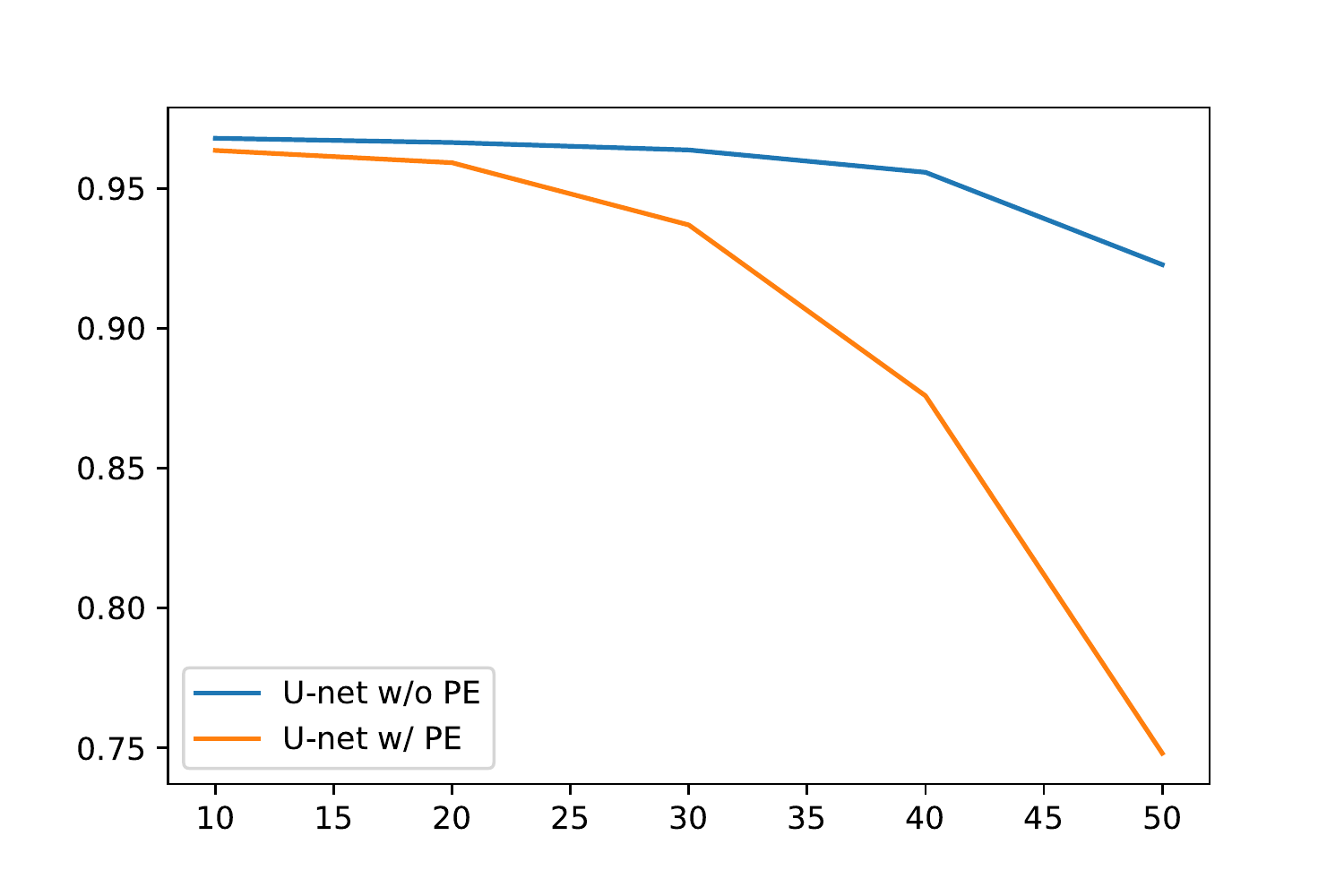}

\vspace*{-10pt}
\caption{Performance for spatially shifted images. Dice coefficient versus the number of shifted pixels. }
\label{fig:jsrt_trans}
\vspace*{-10pt}
\end{wrapfigure}
Moreover, the prior knowledge anchored in the absolute image position is indeed double-edeged sword. 
In fact, as shown in Fig.~\ref{fig:jsrt_trans}, when intentionally shifting each input image by a certain length of pixels in $x$ axis, the U-net model without PE performs better than that with PE. The fact that the model without PE shows inferior performance for large input shifts implies that the model uses position information, which may be provided by padding, to some extent, if not rely much on it.   

\medskip
\noindent
{\bf Effects of different padding methods}~
{\em The padding method does not affect inference for undegraded images but yields difference in the robustness to image degradation. The difference vanishes when PE is employed. } The former implies that zero-padding provides position information more effectively than reflection-padding. The latter provides a further evidence of the above argument--- the U-net model trained on a sufficient amount of data does not need the absolute position in the degradation-free setting.
 
\section{Conclusion}

In this paper, we have discussed how CNNs can utilize absolute image position for segmentation tasks through a series of experimental analyses. In the experiments, we have considered two medical image datasets, JSRT and ACDC. Our conclusion is mixed.

On one hand, the positional encoding, which adds additional channels encoding absolute image positions to the input image so that the CNN will have a direct access to the position information, is effective without doubt. CNNs can thereby memorize the shape of segmentation masks anchored to the absolute image position, leading to improved robustness toward image degradation. We have also found that by mapping the $x$ and $y$ coordinates into the range of $[0,\lambda]$, we can control the relative importance of absolute position to pixel intensity with $\lambda$. 

On the other hand, the usefulness of absolute image position depends on each task and dataset. Our conclusion is that it is less important than our intuition; it may be effective only in limited situations. In fact, the use of absolute position for inference is double-edged sword. Too much denepdency on it will compromise the CNNs' inherent invariance to spatial shift of input images, leading to lower robustness to the image shift. Our experimental results indicate that absolute position information is not essential for accurate inference, even for the two datasets (i.e., JSRT and ACDC) for which  position information must naturally be essential for better inference. This holds true  if there is a sufficient amount of training data and the model has sufficiently capacity and there is no statistical difference  between training and test data.



\bibliographystyle{plain}
\bibliography{library.bib}

\newpage

\section*{Supplementary material}

\setcounter{figure}{5}
\setcounter{table}{2}

\begin{table}[h] \footnotesize
\caption{Architecture of the U-net model used in the experiments.}
\centering
\begin{tabular}{cccccc}
\hline
Layer & Type & Channel & Filter & Stride & Padding \\
\hline\hline
 input &  DataLayer & 1 & - & - & - \\
 conv1\_1 &  Convolution + BatchNormalization & 64 & 3 & 1 & 1\\
 conv1\_2 &  Convolution + BatchNormalization & 64 & 3 & 1 & 1  \\
 pool1 &  MaxPooling & 64 & 2 & 2 & 0  \\
 conv2\_1 &  Convolution + BatchNormalization & 128 & 3 & 1 & 1\\
 conv2\_2 &  Convolution + BatchNormalization & 128 & 3 & 1 & 1  \\
 pool2 &  MaxPooling & 128 & 2 & 2 & 0  \\
 conv3\_1 &  Convolution + BatchNormalization & 256 & 3 & 1 & 1\\
 conv3\_2 &  Convolution + BatchNormalization & 256 & 3 & 1 & 1  \\
 pool3 &  MaxPooling & 256 & 2 & 2 & 0  \\
 conv4\_1 &  Convolution + BatchNormalization & 512 & 3 & 1 & 1\\
 conv4\_2 &  Convolution + BatchNormalization & 512 & 3 & 1 & 1  \\
 pool4 &  MaxPooling & 64 & 2 & 2 & 0  \\
 conv5\_1 &  Convolution + BatchNormalization & 512 & 3 & 1 & 1\\
 conv5\_2 &  Convolution + BatchNormalization & 512 & 3 & 1 & 1  \\
 up1 &  BilinearInterpolation & - & - & - & - \\ 
 cat1 &  Concatenation & - & - & - & - \\ 
 conv6\_1 &  Convolution + BatchNormalization & 256 & 3 & 1 & 1\\
 conv6\_2 &  Convolution + BatchNormalization & 256 & 3 & 1 & 1  \\
 up2 &  BilinearInterpolation & - & - & - & - \\ 
 cat2 &  Concatenation & - & - & - & - \\ 
 conv7\_1 &  Convolution + BatchNormalization & 128 & 3 & 1 & 1\\
 conv7\_2 &  Convolution + BatchNormalization & 128 & 3 & 1 & 1  \\
 up3 &  BilinearInterpolation & - & - & - & - \\ 
 cat3 &  Concatenation & - & - & - & - \\ 
 conv8\_1 &  Convolution + BatchNormalization & 64 & 3 & 1 & 1\\
 conv8\_2 &  Convolution + BatchNormalization & 64 & 3 & 1 & 1  \\
 up4 &  BilinearInterpolation & - & - & - & - \\ 
 cat4 &  Concatenation & - & - & - & - \\ 
 conv9\_1 &  Convolution + BatchNormalization & 64 & 3 & 1 & 1\\
 conv9\_2 &  Convolution + BatchNormalization & 64 & 3 & 1 & 1  \\
 score &  Convolution & 4 & 1 & 1 & 0  \\

\hline
\end{tabular}
\label{tab:arc_Unet}
\end{table}

\begin{table}[h]\footnotesize
\centering
\caption{Architecture of the small-size CNN with four weight layers. }
\begin{tabular}{cccccc}
\hline
Layer & Type & Channel & Filter & Stride & Padding \\
\hline\hline
 input &  DataLayer & 1 & - & - & - \\
 conv1 &  Convolution + BatchNormalization & 64 & 3 & 1 & 1\\
 conv2 &  Convolution + BatchNormalization & 64 & 3 & 1 & 1  \\
 conv3 &  Convolution + BatchNormalization & 64 & 3 & 1 & 1 \\
 score &  Convolution & 4 & 1 & 1 & 0 \\
\hline
\end{tabular}
\label{tab:arc_4layer}
\end{table}

\begin{figure}[h]
\begin{tabular}{cc}
\bmvaHangBox{\includegraphics[width=5.8cm]{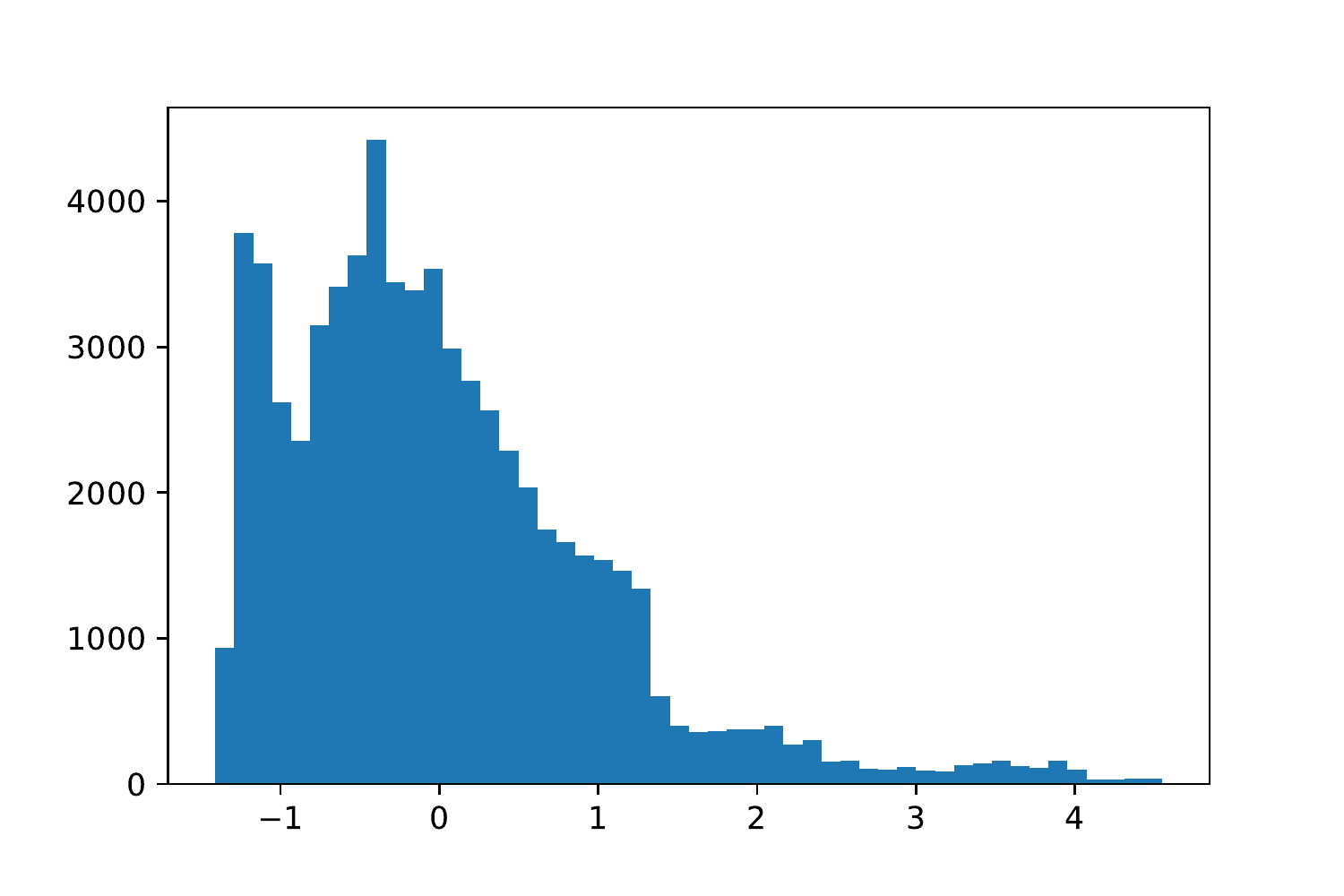}}&
\bmvaHangBox{\includegraphics[width=5.8cm]{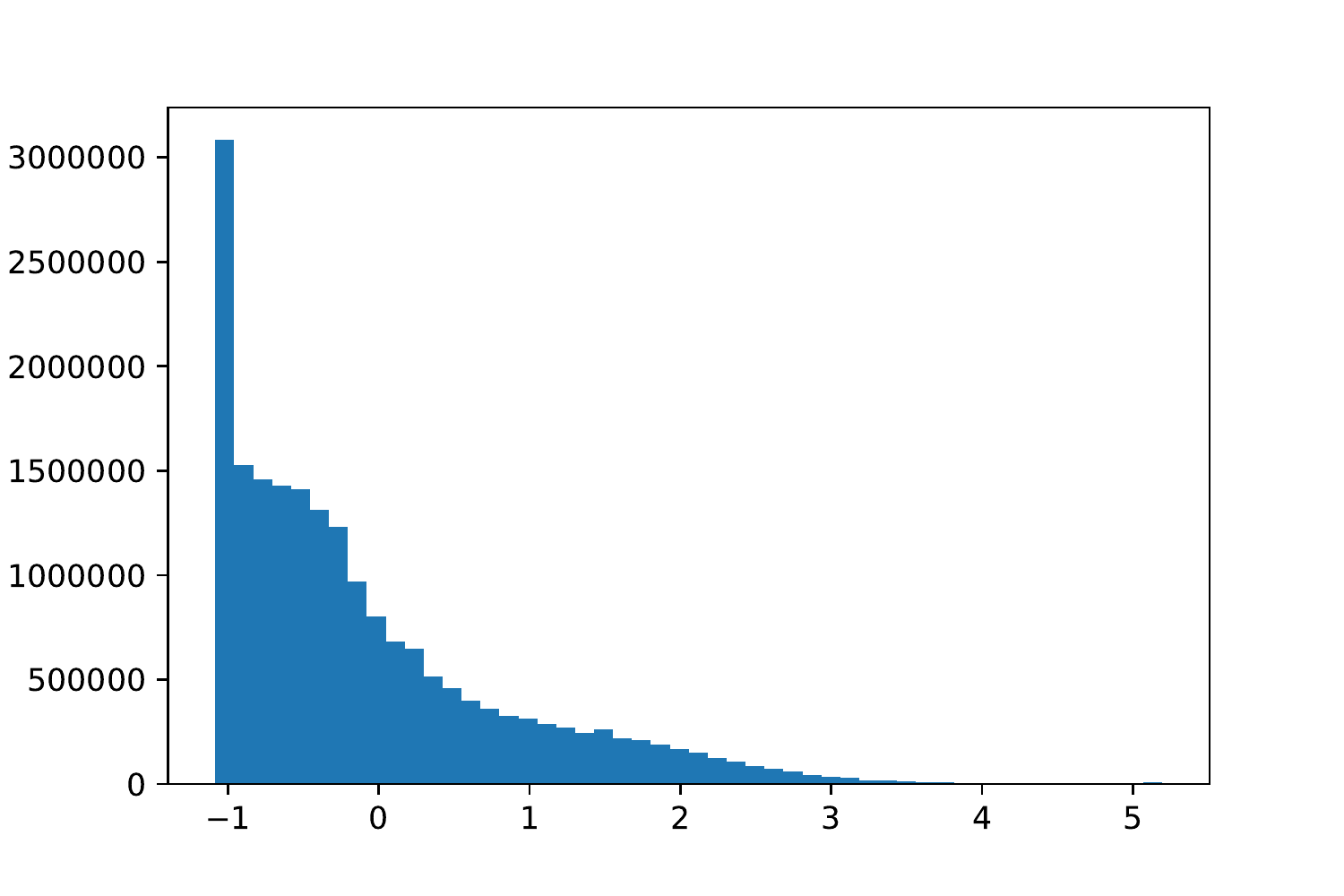}}\\
{\small (a) JSRT}
&
{\small (b) ACDC}
\end{tabular}
\smallskip
\caption{The histogram of pixel intensities of the images of JSRT and ACDC inputted to the CNNs.}
\label{fig:histogram}
\end{figure}

\end{document}